\documentclass{article}
\usepackage[utf8]{inputenc}
\usepackage[english]{babel}
\usepackage{graphicx}
\usepackage{bmpsize}
\usepackage{booktabs}
\usepackage{listings}
\usepackage{wrapfig}
\graphicspath{ {./images/}}
\usepackage{appendix}
\usepackage{subfiles} 
\usepackage{tabularx} 
\usepackage{arydshln}
\usepackage{amsmath}  
\usepackage{graphicx} 
\usepackage[margin=3cm,letterpaper]{geometry} 
\usepackage{xspace}
\usepackage[compress,numbers]{natbib}
\usepackage[final]{hyperref} 
\usepackage[most]{tcolorbox}
\usepackage{subcaption}
\usepackage{textgreek}
\usepackage{enumitem}
\usepackage{fancyvrb}
\usepackage{authblk}

\newcommand{\deltext}[1]{\relax}
\newcommand{\addtext}[1]{#1}

\title{Hybrid CMOS$/$Memristor Circuit Design Methodology}

\author[1]{Sachin Maheshwari}
\author[1]{Spyros Stathopoulos}
\author[1]{Jiaqi Wang}
\author[1]{Alexander Serb,}
\author[1]{Yihan Pan}
\author[2]{Andrea Mifsud}
\author[2,3]{Lieuwe B. Leene}
\author[2]{Jiawei Shen}
\author[2]{Christos Papavassiliou}
\author[2]{Timothy G. Constandinou}
\author[1]{Themistoklis Prodromakis}
\affil[1]{Centre for Electronics Frontiers, University of Southampton, Southampton}
\affil[2]{Department of Electrical \& Electronic Engineering, Imperial College London}
\affil[3]{Novelda AS, Norway, Oslo}

\date{}

\begin{document}
\maketitle
\begin{abstract}
\normalsize
RRAM technology has experienced explosive growth in the last decade, with multiple device structures being developed for a wide range of applications. However, transitioning the technology from the lab into the marketplace requires the development of an accessible and user-friendly design flow, supported by an industry-grade toolchain. In this work, we demonstrate with examples an end-to-end design flow for RRAM-based electronics, from the introduction of a custom RRAM model into our chosen CAD tool to performing layout-versus-schematic and post-layout checks including the RRAM device. We envisage that this step-by-step guide to introducing RRAM into the standard CMOS integrated circuit design flow will be a useful reference document for both device developers who wish to benchmark their technologies and circuit designers who wish to experiment with RRAM-enhanced systems.
 
\end{abstract}
\section{Introduction}

CMOS technology is facing numerous challenges due to the continuous decrease in the device dimension. It is now practically approaching its physical limits of miniaturization, however, due to its negligible static-power dissipation at higher technology nodes, it is still thought to be an important part of future technology. This gradual end of Moore’s law eventually commences a new era in research and development of emerging technologies for future intelligent computing systems. One such emerging technology is called RRAM (also known as memristor), postulated by Leon Chua in 1971  \citep{Chua_1971}, is considered prominent due to its  scalability and in-memory computation capabilities.  

A memristor or memristive device is a two-terminal passive device where the state variable (resistance value) can be altered by allowing electrical current to flow. The peculiarity of the device is that it remembers the resistance value when the power is switched off. In addition, this resistance value can also be increased or decreased depending on the amount of current flow and its direction.
Chua’s theory was overlooked for decades due to the lack of practical applications and the advancement of the technology at that time. Nonetheless, after more than three decades, the theoretical concept of a memristor was proven physically by observing resistive switching in the titanium dioxide device developed at Hewlett-Packard Laboratory  \citep{Strukov_2008}. The primary application of a memristor has been an energy-efficient and scalable memory element, where the data is stored in terms of the resistance value. For high-density memory, the crossbar array architecture \citep{Papandroulidakis_2017} is the most well suited where memristors are connected across each junction of the crossbar. Other unique properties of the memristor are its multi-bit storage capability and state retention with no power applied \citep{Stathopoulos_2017}.

Over the years, memristive devices have been vigorously researched and explored in a wide variety of applications including: non-volatile memory \citep{Papandroulidakis_2017,Stathopoulos_2017,Zha_2016}, programmable logic gates \citep{Gao_2013}, reconfigurable computing \citep{Serb_2018,Papandroulidakis_2018,Dao_2020}, analog computing \citep{Zidan_2017}, neuromorphic computing \citep{Zidan_2019,Laiho2010}, edge-computing \citep{Krestinskaya_2020,Hamdioui_2015}, image processing \citep{Li_2017} and hardware security \citep{Arafin_2015, Jimson_2015}. It has been demonstrated that the use of memristors in the above applications offers an improvement over the state-of-the-art CMOS technology in terms of integration density, energy consumption, multi-level programming capabilities and speed. However, designing memristor-based circuits requires a more realistic model of the memristor's behaviour at physical level in order to attain similar system level design efficiency and accuracy of the much more matured CMOS technology. Thus, the plethora of memristor models has also been proposed in the literature \citep{Prodromakis_2010,Fleck_2016,Vincent_2015,Menzel_2013,Quindeau_2014}. TEAM \citep{Kvatinsky_2013} was a widely and well-established model for in-circuit simulation until recently, where a new model representing a direct link of the experimental data to the model parameter was established \citep{Messaris_2018}.

Memristors have been used both as a standalone circuit (as in logic gates) and also integrated with CMOS for many applications. 
Memristor-based logic design is an emerging concept fuelled by the continually growing need for energy-efficient computation. The memristor property of a variable resistive state due to the applied voltage (amplitude, number of pulses) has lured digital designers to use them for representing ON and OFF logic states as low and high resistance states respectively. Thus, numerous works on memristor-based logic, where the memory operation of the device is combined with the Boolean function have been demonstrated 
 \citep{Kvatinsky_2014,SKvatinsky_2014,Wald_2016,Linn_2010,Levy_2014,Kvatinsky_2012}

The salient feature of memristive devices has gained popularity in crossbar arrays  \citep{Rajendran_2010,Rose_2012} for implementing brain-inspired computing \citep{Gao_2013,Zidan_2019,Rose_2012,Xia_2019}. Here the variable resistance of the memristor is used to mimic the function of a synapse in a neural network. The threshold logic gate based on such a function is configured to implement both the neuromorphic logic and Boolean logic. However, since as a standalone device the memristor suffers from signal degradation and sneak current paths  \citep{Rajendran_2010, Zidan_2013}, it was observed that to implement a large neural network using a threshold function, memristors integrated with CMOS were found suitable.
Most of the memristor-based logic gate approaches are exemplified within the structure of crossbar array such that it performs in-memory computation \citep{Chen_2015}. As demonstrated through the extensive work on circuits, the memristor-based logic design holds great potential for high density and energy-efficient computing. Recently, a hybrid CMOS/memristor chip for neuromorphic computing, has been fabricated with a memristor crossbar array directly integrated with custom-designed CMOS circuits that contain mixed-signal blocks and a digital processor for re-programmable computing \citep{Cai_2019}. This successful integration practically demonstrates efficient hardware implementation in terms of area and energy consumption and also gives an add-on property of re-configurability.

\subsection{Motivation and Challenges}

Memristive devices have a simple and nanoscale structure that consists of Metal-Insulator-Metal (MIM). As a result, fabrication of these devices is similar to the processing of a via between two metal lines. Additionally, due to its unique property of in-memory computation, as well as easy integration with CMOS, it is beneficial in a wide variety of applications. One of the significant benefits of these devices is their multi-bit memory operation capabilities \citep{Stathopoulos_2017}. The other benefit is easy scalability and large-scale parallelism when used within crossbar array structure. However, it is challenging to design high-density memristive memory due to the sneak-current issue which leads to high energy consumption and incorrect read of the resistive state \citep{Rajendran_2010}. Nonetheless, appropriate selector devices have been used to combat this issue to some extent but still requires further innovation in a large-scale implementation. As CMOS technology is not able to provide high performance when it is deeply scaled, due to high leakage current, hybrid CMOS/memristive devices to design novel logic circuits can certainly provide high energy-efficient computations.      

As memristive devices are prominently being researched and are finding their application in the next generation nano-electronics,  it is important to build a step-by-step design methodology for primary researchers to design and validate memristor-based circuits before it is laid onto the wafer or integrated with CMOS. Thus, the circuit designers must work closely with the device engineer and physicist to develop a realistic memristor model or improve further the already developed model such as to include new properties and behaviour characteristics that are essential to designing systems with these devices.   

Although, a common memristive device still does not exist due to their different switching mechanisms and material properties, a design methodology that is versatile, robust, user-friendly and can be integrated with any CMOS technology is the necessity in the current scenario. The purpose of this work is to give a holistic view of the memristor behaviour from experimental analysis to its integration with commercially available CMOS technology. 

\subsection{Structure of the paper}

In this design methodology tutorial, we have used the Cadence Virtuoso design environment for schematic capture, layout design, Verilog-A/Spectre for circuit simulation, and Mentor Graphics Calibre for physical verification including DRC, LVS, and PEX.

The work is organized into various sections and sub-sections. Section 2 provides the behavioural model and the memristor device dynamics that has been developed, characterised and tested within the facilities at the University of Southampton. This section also demonstrates the switching behaviour of the device given an applied pulse/s of a particular width and amplitude. Section 3, introduces the Verilog-A model followed by a step-by-step construction and integration of the device model into the Cadence environment. Consistent with the previous section, this section also presents the switching behaviour using various stimuli for validation. Albeit separated from the main text, the memristor behaviour for the model utilizing exponential fitting \citep{Messaris_2018} is provided in Appendix A. Section 4 focuses on using the RRAM to build circuit blocks for crossbar arrays and to handle the voltages required to interface with the RRAM. Later in this section, the design of a re-configurable gate is demonstrated using RRAM. Section 5, introduces a potential layout of a standard single-cell memristor followed by physical verification using Calibre. Besides, the layout examples in section 4, a memristor array in  a 16 x 16 crossbar structure is also illustrated in this section. The electrical and physical verification in this work has been done using a commercially available $0.18\mu$m CMOS technology.\footnote{The given methodology can be adapted to different technologies and other tool vendors, e.g. Synopsys, Tanner, and or modules e.g. Hspice, Assura, Pyxis, Eldo, etc. The tools specified here are not exhaustive but intend to give you an idea of other possibilities.}

\section{Modelling Memristive Behaviour}
\label{sec:model}

Before attempting to integrate memristors into higher level design we should establish a way of modelling the response of memristive devices for different input stimuli. This section presents a short description of several types of memristive devices from the point of view of a CMOS workflow and then it introduces the concept of a behavioural model for the specific class of devices that are used for the purposes of this work. This phenomenological data-driven model serves as the basis for the rest of the methodology hereof. Of course, given the breadth of available technologies, no model is infallible. Should a different model be more suitable for a given technology, information is presented in a modular manner so as to facilitate an easy transition to a different one.

\subsection{Memristive Devices in a CMOS-oriented Workflow}

Memristors come in different configurations and topologies and considerable work is available in the literature to describe the merits and disadvantages of each of those~\cite{Ambrosi_2019, Im_2020, Zahoor_2020}. It is important to mention that the term \emph{memristor} defines a behaviour rather than a specific form of device and materials. Although the memristor as originally postulated by Chua~\cite{Chua_1976} connects \deltext{magnetic} flux, \addtext{$\Phi$}, and charge, \addtext{$q$}, ($d\Phi = Mdq$) in practical terms it is realised through a variation of an internal state variable (typically resistance) based on the history of biasing that has been applied to the device. That behaviour is what gives rise to the \emph{memory} characteristics typical of such structures~\cite{Chua_2011}.

\deltext{Several different implementations of memristive devices have been put forward} \addtext{Implementations of memristive devices has been demonstrated on a
multitude of material systems} over the past couple decades.
These range from binary metal-oxides~\cite{Brivio_2016}, halcogenides~\cite{Li_2013},
perovskites~\cite{Shi_2013} or even
polymers~\cite{van_de_Burgt_2017} and other 2D
materials~\cite{Younis_2013, Tian_2017}.
Out of these materials metal-oxides are probably the most common, with the original
RRAM-based memristor using TiO$_\text{x}$ as the active layer
material~\cite{Strukov_2008}. Metal-oxide-based RRAM devices are grouped into
two broad categories depending on the mechanism of operation. The first is
\emph{Valence Change Memory} (VCM) the operating principle of which is based on the
modification of the stoichiometry within the active material when external voltage is
applied. This change in stoichiometry is primarily driven by the movement of mobile
oxygen vacancies~\cite{Dittmann_2019}. The second is the \emph{Electrochemical
Metallisation Memory} (ECM) which is based on the formation of a metallic conducting
filament due to partial diffusion of one of the electrodes into the
metal-oxide~\cite{Valov_2011}. \addtext{Metal-oxides are also involved in the
implementation of spin-torque transfer memories~\cite{Fong_2016} that can also
exhibit memristive behaviour~\cite{Lequeux_2016} although in this particular case
transitions between resistive states are based on magnetic coupling rather than
ionic movements.}

For the purposes of this work we are going to focus on a simple \addtext{VCM}
metal-insulator-metal structure based on TiO$_\text{x}$ as an active layer and platinum as top and bottom electrodes. Devices based on the (bottom to top) Pt/TiO$_{x}$/Al$_{x}$O$_{y}$/Pt stack have given exemplary results in the past~\cite{Stathopoulos_2017} and these will serve as the basis for this paper. Of course this approach is not limited to a specific type of device but can be generalised for a broad family of VCM memories.
\addtext{While this manuscript presents a methodology for designing circuits with emerging non-volatile memory technologies and uses an exemplar memristor technology and model, we note that the same principles can be utilized with alternative technologies~\cite{chang2011short} and models ~\cite{giotis2020I} from the literature, including volatile technologies~\cite{Berdan2016}}. 

\begin{figure}
  \centering%
  \includegraphics[width=0.45\textwidth]{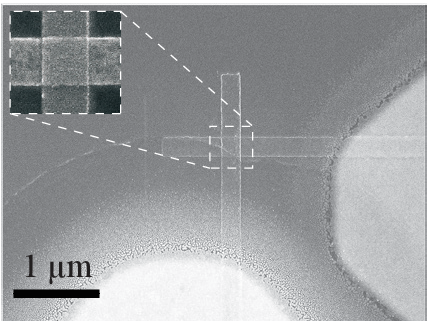}
  \includegraphics[width=0.5\textwidth]{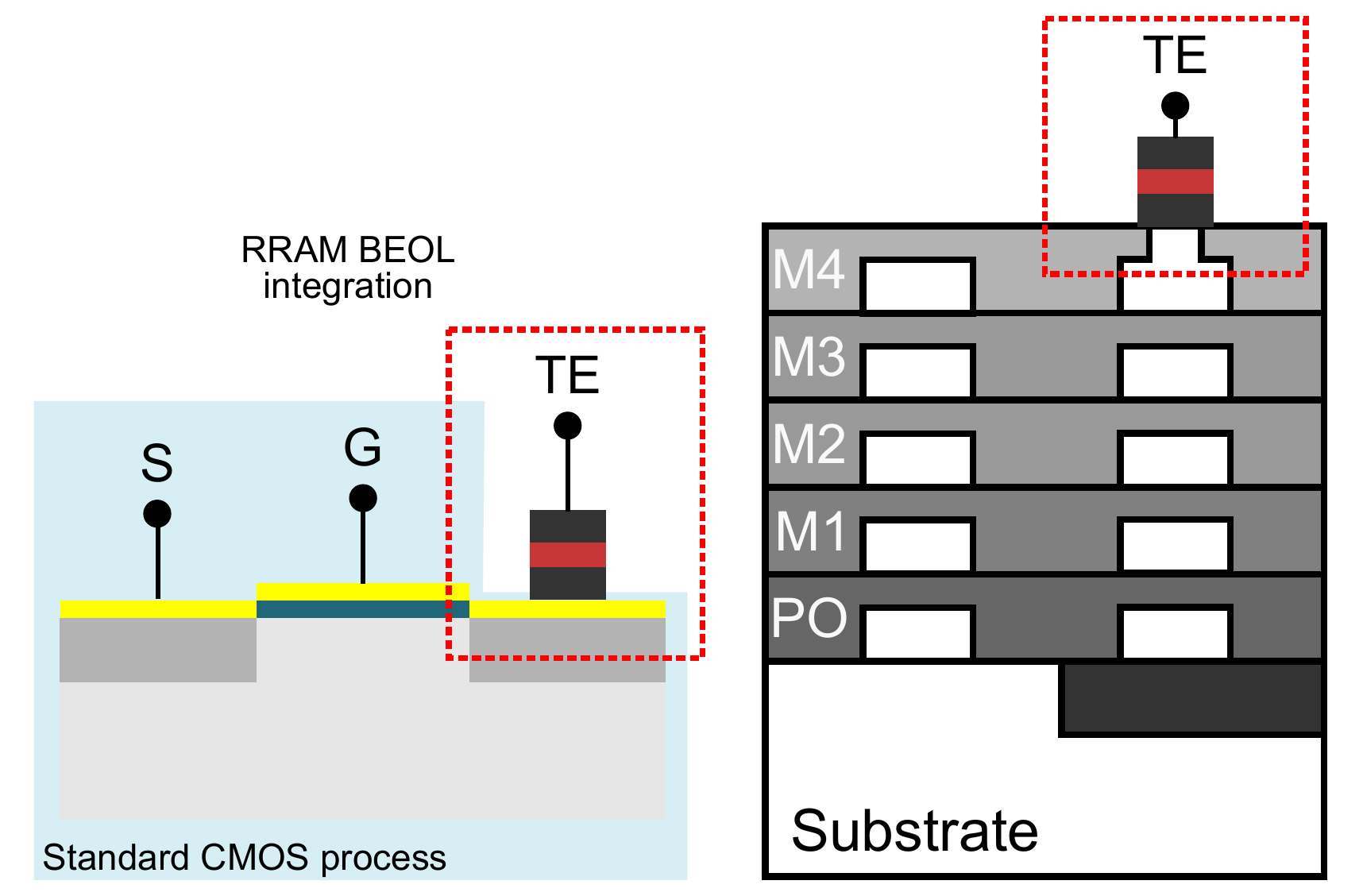}
  \caption{\label{fig:devices}Photo of a crosspoint sub-1$\mu$m$^\text{2}$ active area of a single device (top). Example of possible back-end-of-line (BEOL) integration on a standard CMOS process. The simplicity of the multi-layer structure allows for straightforward post-CMOS integration (bottom left and right). One of the terminals of the device is the biasing electrode used to program the device while the other is connected to the CMOS drain. Toggling the gates of the transistor effectively acts as a selector for that specified device.}
\end{figure}
In its simplest form a typical VCM memory is a stack of active material, usually a dielectric, between two metal electrodes (fig.~\ref{fig:devices}). A variety of metal combinations can lead to different contact behaviours as discussed previously~\cite{Michalas_2018}. The relative simplicity of the structure is important for the mask-efficient integration process of such device in a CMOS standard process.

A CMOS integration effort featuring memristors require practical, fast and accurate models for the technology at hand. Depending on the underpinning physical mechanism, different models have emerged for the specific family of memristive materials and devices, be it resistive valence change memories~\cite{Fleck_2016}, spin-torque transfer memories~\cite{Vincent_2015} or electrochemical metallisation memories~\cite{Menzel_2013}  to name a few. As it is probably not possible to have a model for every existing combination of materials from an integration point of view, what is more important is the overall \emph{behavioural} characteristics of the device and how these are affected by volatility, noise and thermal effects. After all, Chua's definition of a \addtext{generic} voltage-driven \deltext{memristive response} \addtext{memristor} can be pinpointed to a set of two basic equations~\cite{Chua_1976}.
\begin{align}
               v &= R(x)i\\
  \dfrac{dx}{dt} &= f(x, v)
\end{align}
where $v$, $i$ and $R$ the voltage, current and resistance of the device,
respectively, which can be dependent upon an internal \deltext{system-wide} state
variable, $x$. \addtext{This is what Chua calls the differential algebraic form of a memristive response~\cite{Chua_2015_everything}}.

\subsection{Phenomenological Models and Device Dynamics}

In a phenomenological model there are many approaches to model the temporal
evolution of the function $R(x)$. Kvatinsky et al.~\cite{Kvatinsky2015} proposed a linear/exponential expression where others~\cite{Yakopcic2013} have put forward a hyperbolic sinusoidal function which is appropriate for metal-insulator-metal devices. In the simplest approach, the function $f(x,v)$ is approximated by a product of decoupled equations in the form of $f(x,v) = s(v)\cdot g(x)$. Function $s(v)$ describes the voltage sensitivity of the device while $g(x)$ is the \emph{window function} that delimits the operational boundaries of the device. This form of decoupling is necessary in order to address boundary issues with the original proposition of the memristor model as presented by Strukov et al.~\cite{Strukov_2008}. Choosing a window function that is generalised enough to be used across different devices and technologies has been a challenging effort and several have been presented that have been tailored to specific modelling efforts~\cite{Kvatinsky2015, Prodromakis_2011}.

For the purposes of this methodology we will be relying on the phenomenological model presented in~\cite{messaris2017tio2} as it has been used extensively by the authors before for the devices described in this section. It is also generic enough for voltage-modulated devices \addtext{and is capable of emulating the transient response during switching depending on the time step resolution.} The state variable in this case is the resistive state itself, the sensitivity function is a voltage-dependent exponential function while the window function employed is a state-dependent \deltext{exponential} \addtext{quadratic}. \addtext{Of course the internal state variable itself will depend on a series of additional physical parameters, such as temperature. How this affects the resistive response of the device is being addressed elsewhere~\cite{Dhirendra2021A} as it
goes beyond the scope of the present work}. Being state and voltage dependent a \emph{behaviour} of the device can be predicted for a given voltage stimulus at a specific resistive state.

\begin{align}
  i(R, v) &= \left\{\begin{array}{ll}
    a_p(1/R)\sinh{(b_p v)}  & v \ge 0 \\
    a_n(1/R)\sinh{(b_n v)}  & v < 0
  \end{array}\right.
  \label{eq:model_ifunc}
\end{align}  

\begin{align}
    \dfrac{dR}{dt} &= g(R, v) = s(v) \cdot f(R, v)
\end{align}
with $s(v)$ being the switching sensitivity function\addtext{, describing the switching rate changes with voltage amplitude.}
\begin{equation}
  \label{eq:model_sfunc}
  s(v) = \left\{\begin{array}{ll}
      A_p(-1+\exp{(t_p|v|)}) & v > 0 \\
      A_n(-1+\exp{(t_n|v|)}) & v < 0 \\
    0 & \text{\deltext{otherwise}\addtext{$v=0$}}
  \end{array}\right.
\end{equation}
and $f(R,v)$ the window function. \addtext{Parameters $r_{p,n}$ depend on the voltage in a polynomial fashion and describe the boundaries of the state variable.}

\begin{equation}
  \label{eq:model_ffunc}
  f(R,v) = \left\{
    \begin{array}{ll}
      (r_p(v)-R)^2 & v > 0 \\
      (R-r_n(v))^2 & v < 0 \\
      \phantom{-}0 & \text{\deltext{otherwise}\addtext{$v=0$}}
    \end{array}
  \right.
\end{equation}

\begin{figure}[ht!]
  \centering%
  \includegraphics[width=0.65\textwidth]{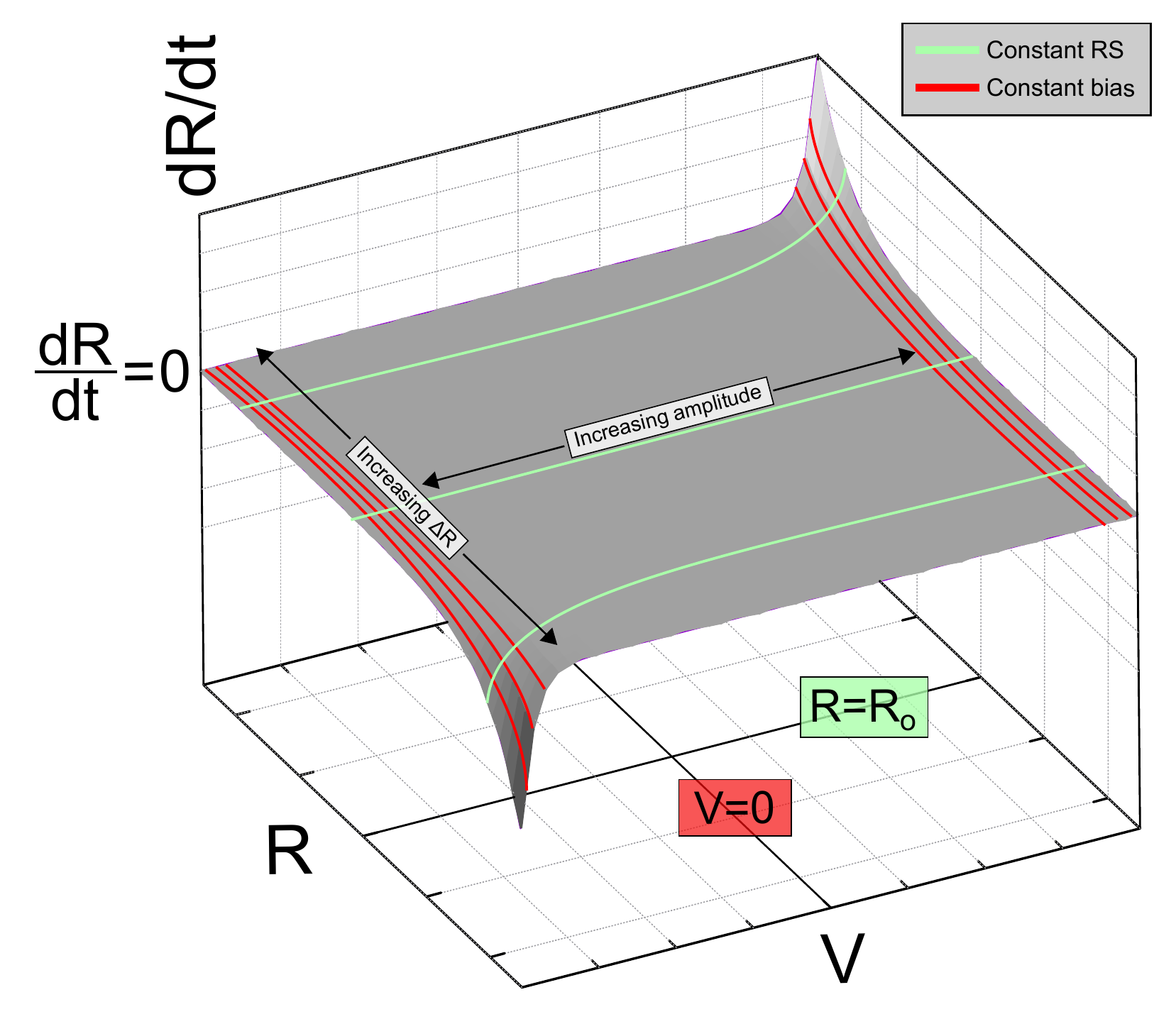}
  \caption{\label{fig:surf}Switching surface based on the model used for this
  work (reproduced from~\cite{Messaris_2018}). \addtext{RS in this context stands
  for ``resistive state''.}}
\end{figure}

\deltext{Parameters $r_{p,n}$ depend on the voltage in a polynomial fashion and describe the boundaries of the state variable.} All other variables are free fitting variables. Fig.~\ref{fig:surf} depicts an indicative surface as described by the equations~\ref{eq:model_ifunc},~\ref{eq:model_sfunc} and~\ref{eq:model_ffunc}. The surface plots the switching rate from equation~\ref{eq:model_ifunc} as a function of the applied stimulus and current resistive state (ie.\ the state variable).

\addtext{In the particular case illustrated in Fig.~\ref{fig:surf}, there is a region around both $R_0$ (the initial resistance, typically in the middle of the operating range of the device}) and \deltext{$V=0$} \addtext{$v=0$} where voltage stimulus does not produce any appreciative change in the state variable. As the offset from the initial resistive state increases the effect of amplitude gets increasingly pronounced indicating a typical bipolar memristive behaviour.
\begin{figure}[ht!]
  \centering%
  \includegraphics[width=0.7\textwidth]{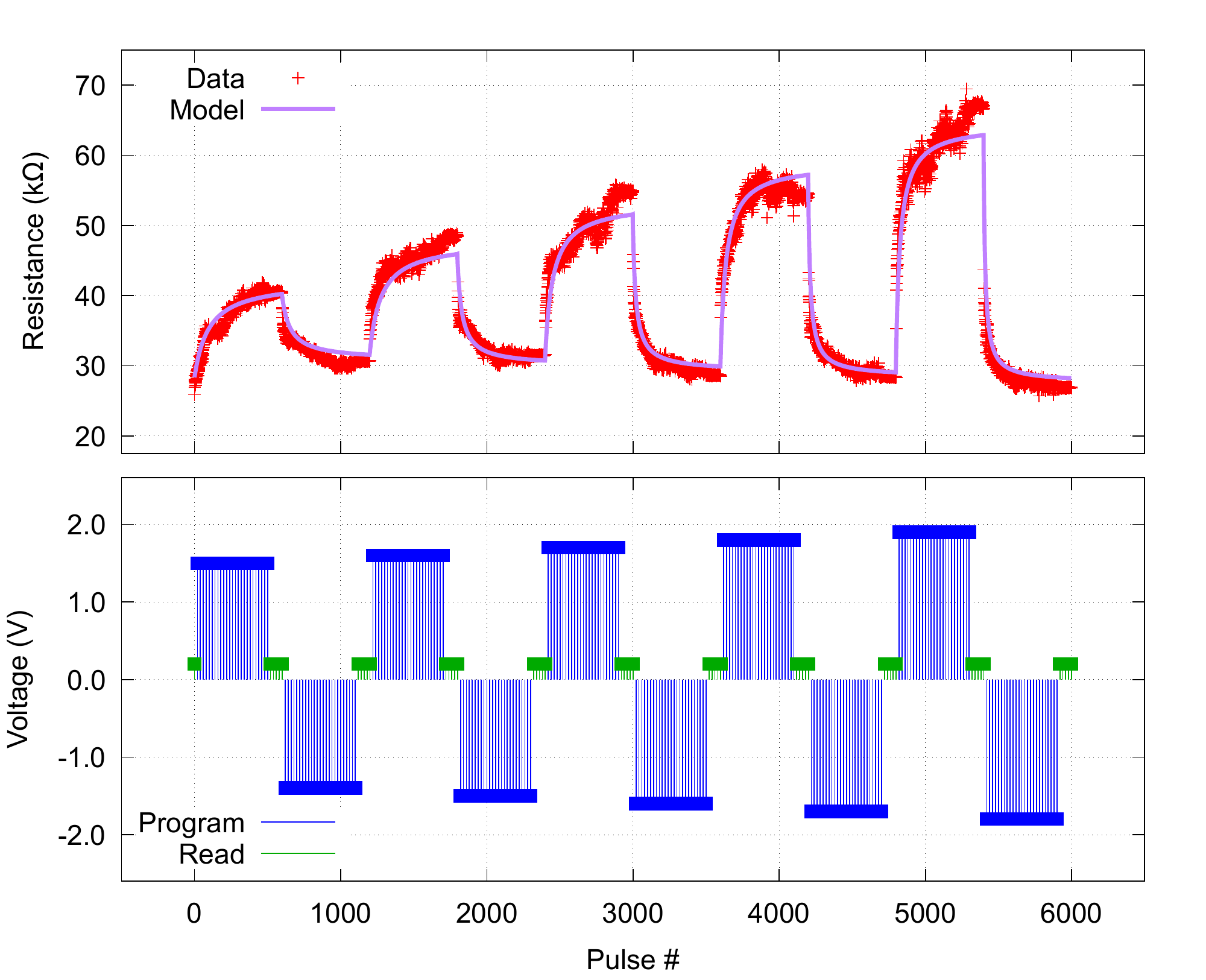}
  \caption{\label{fig:fitted}Analogue switching (top) of a
  Pt/Al$_\text{2}$O$_\text{3}$/TiO$_\text{2}$/Pt RRAM device with respect to an applied stimulus (bottom). Read-outs (ie. samplings of the resistive state) are interspersed between programming phases. The modelled stimulus for the same input is noted with a solid coloured line (top). A number of 500 programming pulses were used to elicit this kind of response from the device. A set of 100 reading pulses was added between programming phases to assert stability of the current resistive state.}
\end{figure}

\begin{figure}[ht!]
  \centering%
  \includegraphics[width=0.65\textwidth]{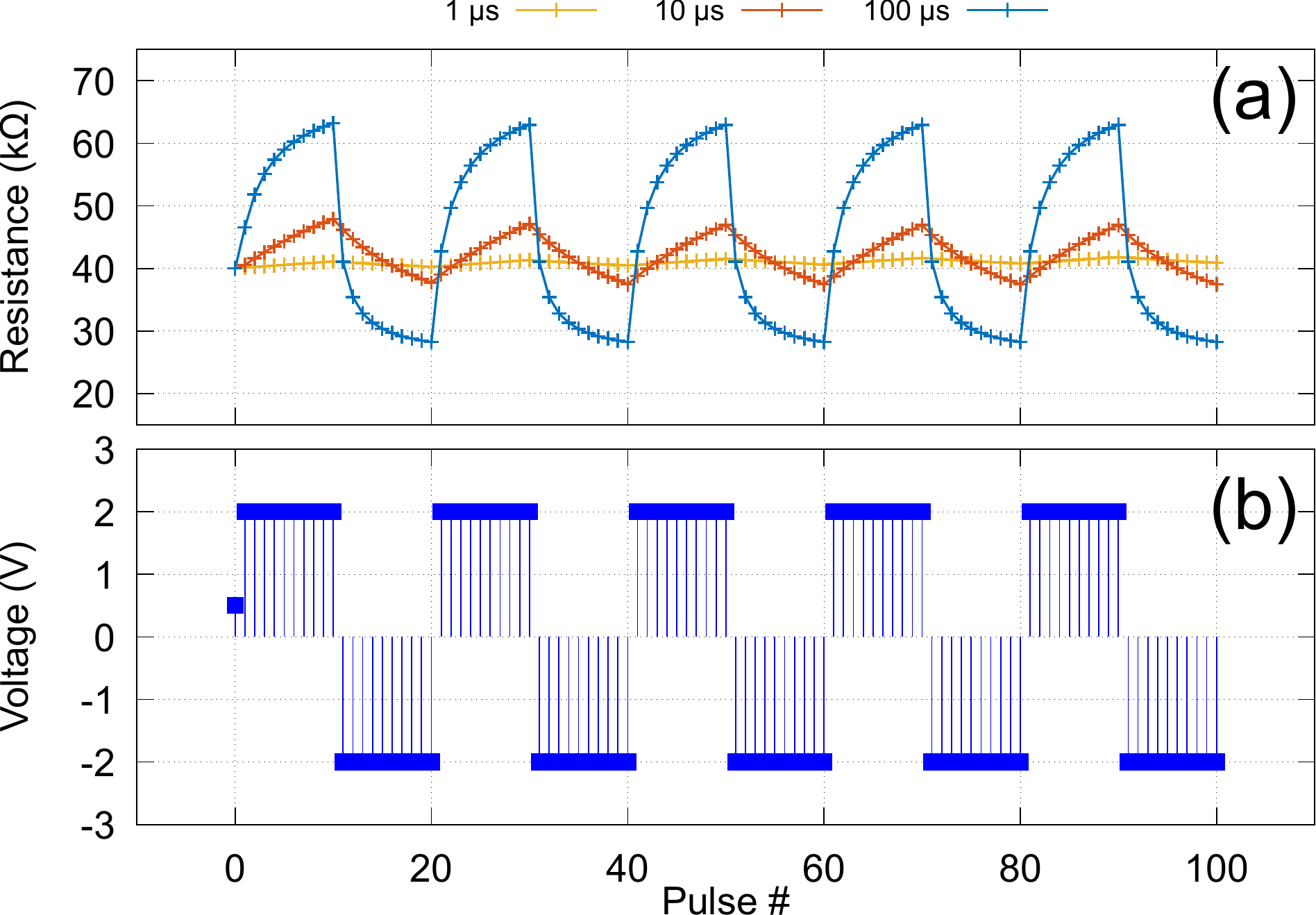}
    \includegraphics[width=0.65\textwidth]{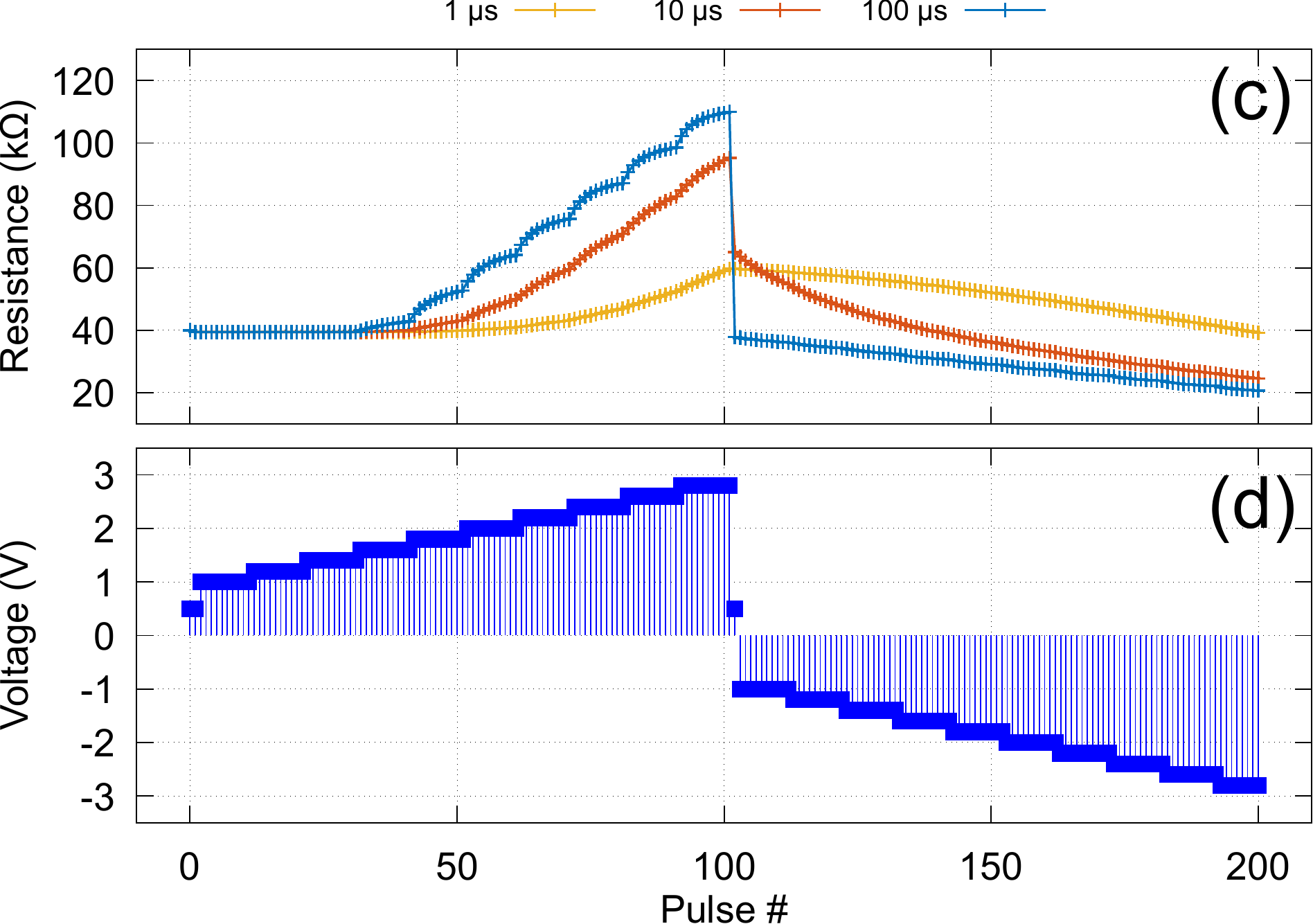}
  \caption{\label{fig:mod-dep} Behavioural model response based on the \addtext{response of the} device shown in fig.~\ref{fig:fitted}. (top) Dependence of the modelled response on applied pulse width for constant amplitude pulses and (bottom) for programming voltage ramps. \deltext{As in the parent device read-out at 0.5~V elicits no change in the RS.} \deltext{Different shades represent increasing programming pulse width.} Batches of 10 programming \addtext{and read-out} pulses were used throughout.
  \addtext{Traces (a, c) indicate the response of the modelled device for two different input waveforms (b, d) at three different pulse widths (1, 10 and 100~$\mu$s).}}
\end{figure}

A response of a bipolar device to an increasing amplitude voltage stimulus can be seen in Fig.~\ref{fig:fitted}. In this case positive pulses cause an increase in the resistance of the device whereas negative bias does the opposite. The device exhibits a gradual rate of increase (or decrease) to the given stimulus which is captured accurately from the model (solid coloured line on Fig.~\ref{fig:fitted}) as described by the equations above. \addtext{Based on the captured model,
fig.~\ref{fig:mod-dep} displays simulated device responses based on fixed amplitude (top) and
varied amplitude (bottom) stimulations for three different pulse widths (1--100~$\mu$s).}
\addtext{Whilst a typical voltage source is assumed throughout these examples, we note that this methodology extends to include alternative sources to reflect the need of distinct input waveforms as required by different applications}.

Translation of the model into a more systems-specific Verilog-A code can be done directly as discussed in~\cite{Messaris_2018} using the domain integrator operator \texttt{idt()}. To avoid the discontinuities of the piecewise function these can be reshaped using the sigmoid function
\begin{equation}
  \theta(x) = \left({1 + \exp{\left({-\frac{x}{b}}\right)}}\right)^{-1}
  \label{eq:sigmoid}
\end{equation}
where $b$ is a hardcoded parameter depending on the function that equation~\ref{eq:sigmoid} is used to shape. For equation~\ref{eq:model_sfunc}, $b=10^{-6}$ whereas for equation~\ref{eq:model_ffunc}, $b=10^{-3}$. Finally to account for the steepness of the exponential in equation~\ref{eq:sigmoid} the \texttt{limexp()} operator is used to bound numerical overflows.

Accurately capturing the transient behaviour of the device under a specific set of stimuli is definitely the primary characteristic of a device model. However there are also issues that need to be considered such as volatile effects~\cite{Gupta_2017}, thermal static characteristics, noise at rest~\cite{Puglisi_2018} or during programming~\cite{Stathopoulos_2019} as well as variability issues. \addtext{Above all, an important aspect of the process lies in assessing the parasitic elements involved, typically parasitic resistances and capacitances. Early in the design cycle, those can be estimated based on fundamental equations of conductance/capacitance and knowledge of device geometry at the device level (initial values to directly plug into the behavioural device model) 
~\cite{SerbA2015},~\cite{Chen2013}}. All of these add additional requirements to device models if they are to emulate all facets of the device response.
Integration of the device model, in a TCAD workflow requires certain considerations in order to accommodate the discrete nature of a typical simulation scenario. Providing a device does not exhibit any appreciable drift resulting in unexpected volatility, an arbitrary programming pulse of duration $T$ can be adequately approximated as a short pulse sequence corresponding to the discrete time step of the simulator, $t_s$. In effect this process discretises the input waveform.

\begin{equation}
  T = \sum_{i=0}^{k} t_s
  \label{eq:mod_discr}
\end{equation}
This method ensures that the total amount of energy used to actuate the device remains the same regardless of the way the pulse is applied.

\section{Integration of Memristor Model into Cadence}

After stating the behaviour and the characteristics of the physical model in the previous section, the Verilog-A memristor model, representing the behaviour of memristive devices through physical equations, will be explicated in this section. Then, this work will go through the detailed electrical design process under a 0.18um CMOS technology that integrates the proposed Verilog-A model into a ASIC design work flow, in this case Cadence Virtuoso using Spectre as a simulator. The simulation applies determined pulses to measure switching behaviours of proposed model, followed by result analysis that contains calibration and validation. This is then built upon in the next section to obtain a hybrid CMOS-Memristor cell design.


\subsection{Verilog-A memristor model}\label{subsec:veriloga}
Verilog-A, a hardware description language, is widely used in semiconductor industry due to the simplicity and flexibility in executing it on circuitry simulators, including Spectre, HSPICE, Eldo, ADM and others \cite{1393990}. It can be utilised to represent the behaviour of memristive devices through physical equations, laying the foundations to enable inclusion of these devices into integrated circuits. This subsection focuses on the specific Verilog-A memristor model that uses quadratic fitting as proposed in \cite{messaris2017tio2}. We will go through the model in terms of the main concept of derivation, and the processing mode which makes it suitable for fast and large-scale simulations. The Verilog-A code, especially for the procedure and significant parameters, will be explained before providing users with approaches and restrictions when applying the proposed model. Besides, another model representing the same device that using exponential fitting in \cite{Messaris_2018} will be presented in Appendix, where the $RS$ range is specific to [$4.5k\Omega$, $6.0k\Omega$] and [$10k\Omega$, $17k\Omega$] respectively. 

As the differential algebraic equation (DAE) set of the physical model (Eq.\ref{eq:model_ifunc}-\ref{eq:model_ffunc}) has been explained in the previous section, the behaviour of the device can be obtained by applying a stimulus at a specific resistive state. Besides, the `absolute threshold' function in Eq.\ref{eq:Rv} shows that the applied bias voltage sets the threshold/boundary of the $RS$.
It inspired us to convert the DAE set to RS time-response equations analytically under constant bias voltage as shown in Eq.\ref{eq:model_analytical}. Parameters in these equations are fully provided in \cite{messaris2017tio2}. The operation of the Verilog-A model presenting the $I-V$ characteristic will be divided into two steps which will be repeated multiple times: 1) programme and keep tracking the last $RS$ from initial $RS$ under the applied voltage stimulus according to Eq.\ref{eq:model_analytical}; 2) update the current flowing through the device based on the last $RS$ through Eq.\ref{eq:model_ifunc}. The concept of breaking the DAE into two parts in Verilog-A model avoids integration of two variables: $RS$ and bias voltage, which speeds up its execution in large-scale simulations.

\begin{equation}
    \label{eq:Rv}
  r(v) = \left\{
    \begin{array}{ll}
        r_p(v) = r_{p,0}+r_{p,1}v,  & v > 0 \\
        r_n(v) = r_{n,0}+r_{n,1}v,  & v \leq 0  \\
    \end{array}
  \right.
\end{equation}
where $r_{p,0}$, $r_{p,1}$, $r_{n,0}$ and $r_{n,1}$ are fitting parameters extracted from physical device. This equation interprets the $RS$ boundaries depending on the bias voltage.

\begin{equation}
  \label{eq:model_analytical}
  R(t)|_{V_b} = \left\{
    \begin{aligned}
        \frac{R_0+s_p(V_b)r_p(V_b)(r_p(V_b)-R_0)t}{1+s_p(V_b)(r_p(V_b)-R_0)t} \quad & for\quad V_b > 0 \& R < r_p(V_b) \\
      \frac{R_0+s_n(V_b)r_n(V_b)(r_n(V_b)-R_0)t}{1+s_n(V_b)(r_n(V_b)-R_0)t} \quad & for\quad V_b \leq 0 \& R > r_n(V_b) \\
      R_0 \quad  \quad \quad \quad \quad \quad \quad \quad & \text{else}
    \end{aligned}
  \right.
\end{equation}
This equation illustrates that the initial $RS$ ($R_0$) changes dependent on the bias voltage ($V_b$) in a fixed pulse duration ($t$), with the combination of switching sensitivity (Eq.\ref{eq:model_sfunc}) and window function (Eq.\ref{eq:Rv}).

With the aid of equations described in \cite{messaris2017tio2}, our Verilog-A memristor model (for the in-house fabricated $Pt/TiOx/Pt$ device) was implemented as presented in Listing 1. The Verilog-A model is the simple and understandable combination of the equations in section \ref{sec:model}. The structure is organised as follows:
\\
\\
\\
\\
\\
\\
\\
\\
\begin{wrapfigure}[41]{r}{0.5\textwidth}
  \centering
    \includegraphics[width=0.5\textwidth]{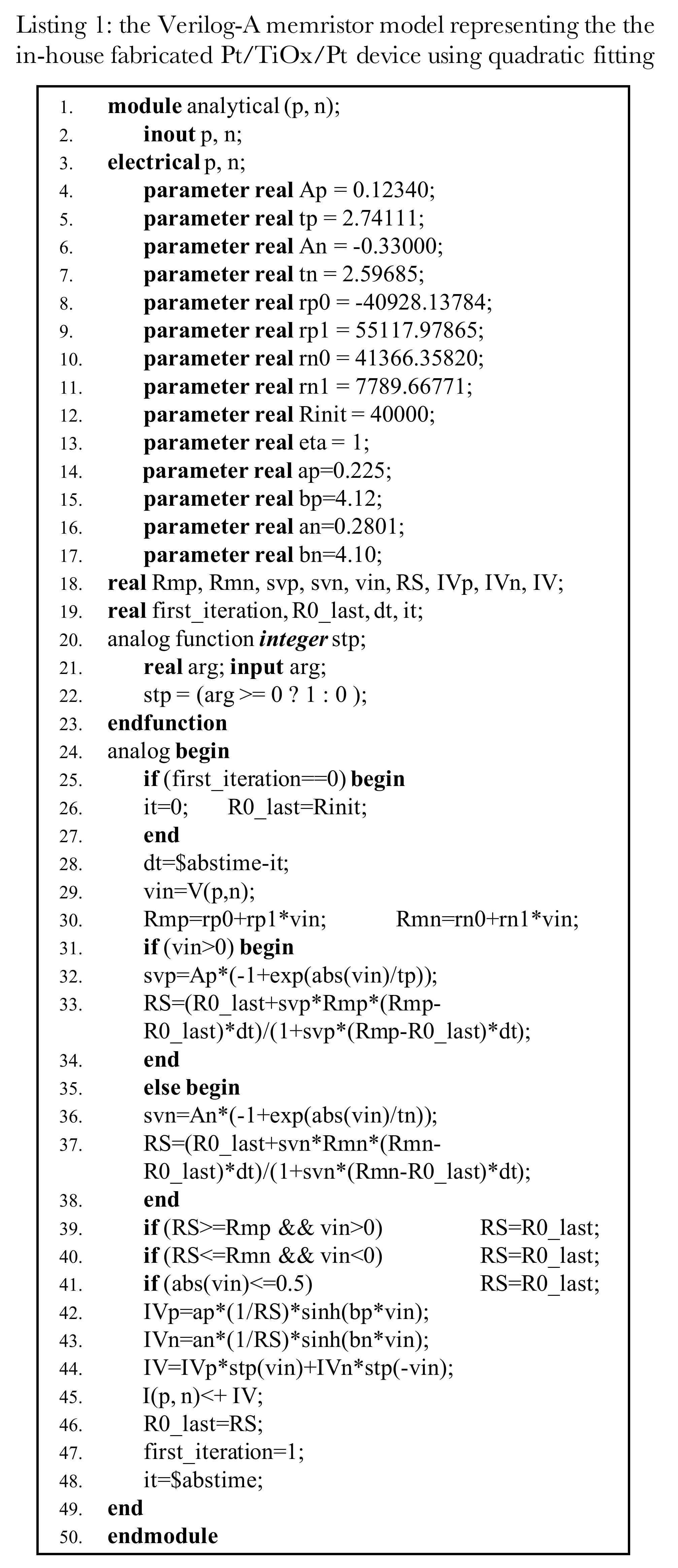}
    \vspace{-20pt}
\end{wrapfigure}
\begin{itemize}
    \item Defining $p,n$ as `inout' ports, from where the bias voltage can be calculated and the current flowing through ports, will be the output. The transient RS can be obtained (from lines 1-3, 29, 42-45), by dividing the applied voltage and the current across the device.
    \item Defining the fitting parameters (switching sensitivity, window function and I-V relationship parameters), switching direction parameter and initial RS (lines 4 to 17). Among these, fitting parameters are extracted from the experimental results whilst using our in-house fabricated device by applying multiple voltage levels. More details about the extraction can be found in \cite{messaris2017tio2}. The switching direction parameter, $eta$, can be defined as 1 or -1, depending on the desirable direction. The initial RS can be set within a proper RS range that is determined by the threshold function (line 30). 
    \item In Eq.\ref{eq:model_ifunc}, two branches of current were derived depending on the polarity of the bias voltage (positive/negative), however, two branches of current will be calculated without recognising the polarity of bias voltage (lines 42-43). Therefore, a `step function' was defined as a multiplexer (lines 20 to 23) in order to choose the proper current branch as output in line 44. When `Vin' is positive, the function `stp(Vin)' is one and the current branch `IVp' will be output, whereas negative bias voltage induces opposite results.
    \item Lines 25 to 27, assign the parameter initial RS to the latest RS that can be used for calculation in the first iteration. The time step is obtained from line 28 by subtracting the absolute time from the reference time to help track each time-step duration. After one calculation run, the reference time will be assigned by the previous absolute time in line 48.
\end{itemize}

\begin{itemize}
    \item The boundaries of RS dependent on the bias voltage based on the chosen window function is calculated in line 30, according to Eq.\ref{eq:model_ffunc}. Line 41 sets the constraint, that below $0.5V$ the bias voltage will not induce switching which has been set as a read voltage (details in 3.2.2).
    \item Lines 31 to 41 calculates RS under the constant bias voltage condition, including situations when the operation is within or exceeds the boundaries of the window function.
    \item After deriving the RS at a specific voltage, the current will be updated (Eq.\ref{eq:model_ifunc}) and passed to ports (line 45). Finally, the initial RS will be updated for the next iteration (lines 46 to 48).
\end{itemize}


This is the data-driven model that we obtained by applying multiple voltages on the device based on the parameter extraction algorithm, of which more information  can be found in \cite{messaris2017tio2}. In this stage, we present $RS$ ranges that have been proved to fit our physical model with a low RMS error. Therefore, users are supposed to distinct the operational $RS$ range, which help them set the initial $RS$ ($R_{init}$ in Listing 1, line 12). Besides, the restriction of $RS$ range limits the applied bias voltage. For positive voltage, only when satisfying both $Vin>0$ and $RS<Rmp$ can induce switching. Combining lines 8, 9, and 30-34, the positive voltage that is above $1.4V$ at least can trigger switching in this model. Users are suggested to do a rough calculation before setting the initial $RS$ and bias voltage. If users set these parameters beyond the `window function' (Eq.\ref{eq:Rv}), the simulation can still be completed but keep the $RS$ as initial setting stage.

Once familiar with the design concept and operation of proposed Verilog-A memristor model, the users can utilise it to 1) conduct static current-voltage measurement at specific RS within boundaries of the state variable; 2) gather transient switching characterisation by applying different bias voltages on defined initial $RS$s. In order to process the above utilisation, two stages should be set up: 1) read $RS$ through applying $0.5V$ triangular pulse, which prevent the device from switching; 2) write, or change the $RS$, by applying pulses with defined duration/width, amplitude, polarity and numbers of pulses. Fitting parameters can also be changed to represent other memristive devices. At this stage, the model does not incorporate AC analysis/small signal modelling, noise performance, parasitics and device variation. 


\subsection{Process flow for electrical design}
\begin{figure}[ht!]
\centering
\includegraphics [width=14cm]{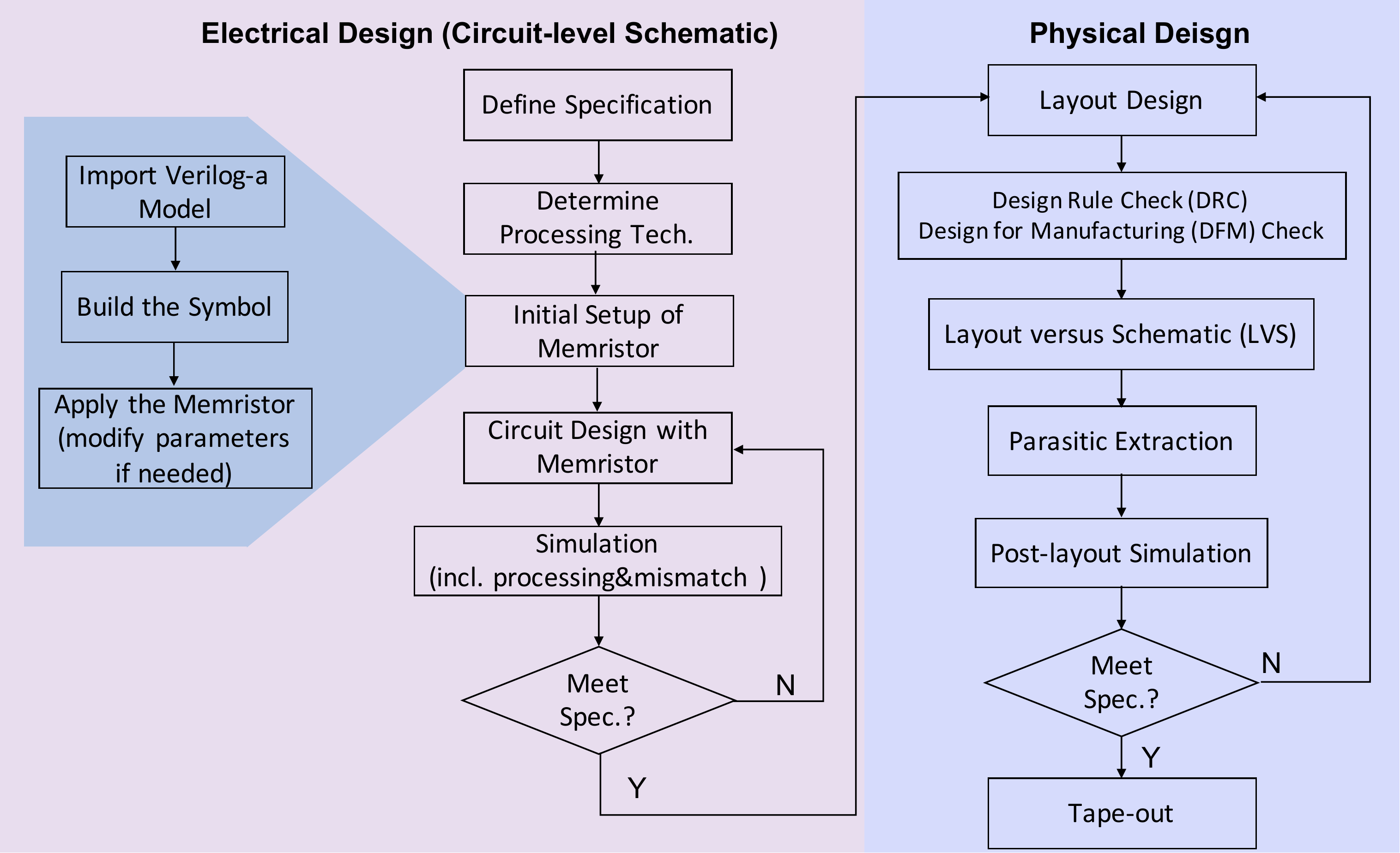}
\caption{Design Flow for analogue and mixed-signal systems. The design flow is divided into two parts: electrical and physical designs, while the model integration is in blue block.}
\label{ProcessFlow}
\end{figure}

The analogue and mixed-signal system design flow is presented in Figure \ref{ProcessFlow} including both electrical and physical design steps. The former will be described into more detail in this section, whilst physical design is presented later in section 5.

\subsubsection{Import verilog-A model and build the symbol}
\begin{figure}[ht!]
    \centering
    \includegraphics [width=14cm]{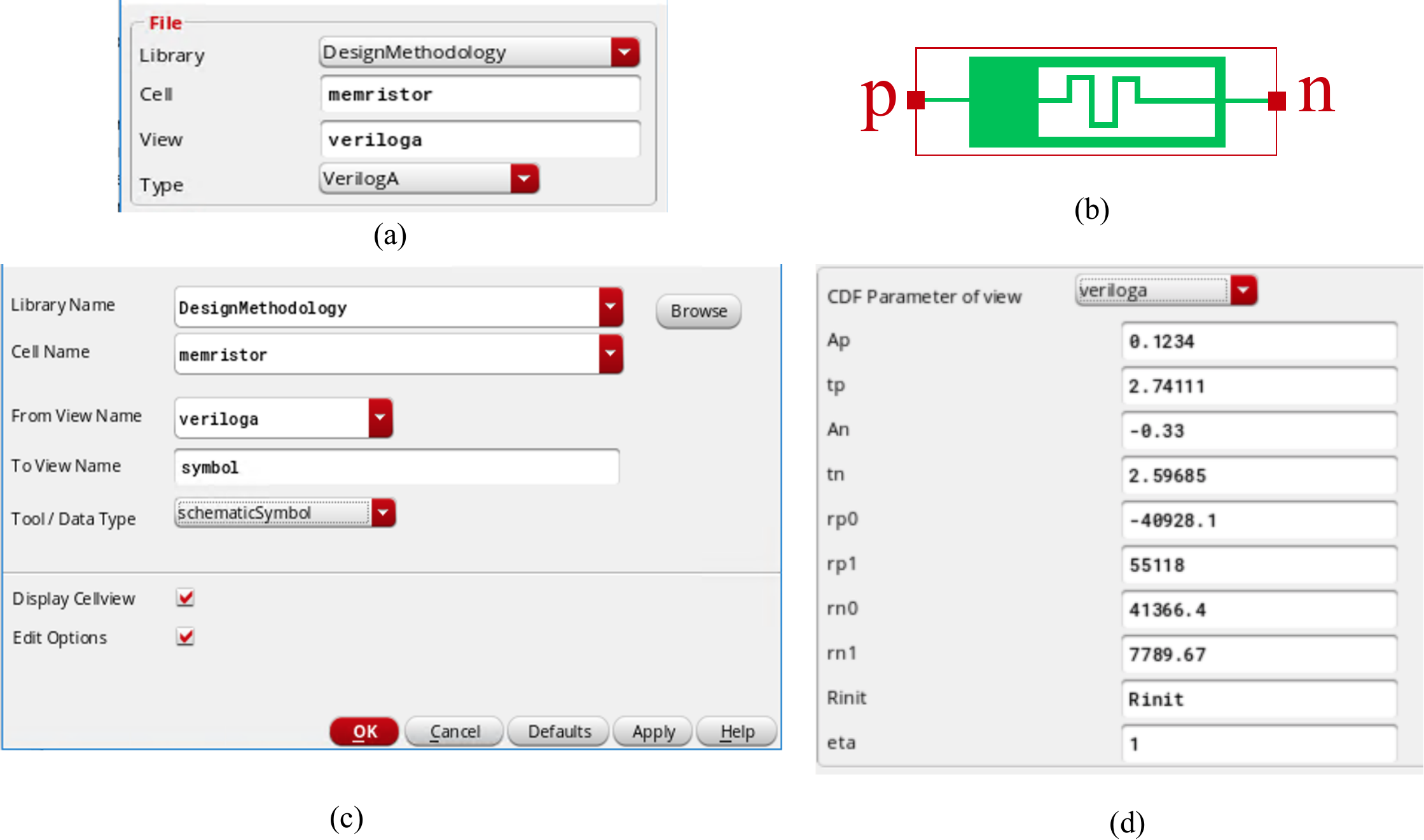}
    \caption{Operation sequence of importing memristor into Cadence. (a) Process of building a new cell view in Verilog-A type for memristor. (b) Symbol of memristor. (c) Process of creating symbol for memristor from Verilog-A cell view. (d) Parameters of memristor can be modified in the object properties window.}
    \label{Steps}
    \end{figure}

In order to utilise the device in schematic design as well as in simulation, a symbol was created that links to the Verilog-A model. This enables the device to be added in a schematic for a design, or for a testbench to be simulated. The instructions to enable this are as follows:
    \begin{itemize}
        \item Creating a library and refer to the chosen technology. Before integrating the memristor with CMOS, users are supposed to be familiar with the behaviour of the memristor quantitatively such as the range of high/low resistive state, the allowed range of applied voltage/current, static I-V characteristic, etc. This allows the user to choose a more suitable memristor technology, as well as define circuit performance. A quantitative analysis based on 0.18um CMOS technology is provided in section 3.3 to provide users with a template whilst using the operational range of our device, as well as measurement methodology. In this work, the Verilog-A model was integrated into Cadence and 0.18um technology with four fabrication metal masks was chosen for demonstration. The proposed memristor was fabricated with Metal4 as ports..\par
        \item A cell view is created in a specific library with Verilog-A type, named `memristor' in library `DesignMethodology' (Figure \ref{Steps}(a)) where Verilog-A code is scripted.
        \item From the toolbar in a text editor, a symbol is created from the Verilog-A memristor model (Figure \ref{Steps}(c)).
        \item The position of ports are assigned at left and right sides which can be automatically detected from the Verilog-A code, followed by the symbol design (Figure \ref{Steps}(b)).
    \end{itemize}
    
\subsubsection{Integrate model and setup simulation}
After creating the symbol, the memritive device is ready to be applied into circuit visually, whose parameters can be changed to represent other models (in Figure \ref{Steps}(d)). However, a read voltage below threshold is required to measure the $RS$ of memristor indirectly. Thus, to trace the changing $RS$ appropriately, the recommended testbench and stimulus is given in the following instruction.

\begin{itemize}
    \item The memristor model can be applied into a schematic, operating with other devices including transistors, resistors and capacitors. Design examples with detailed design information, including primitive cells for controlling the memristor and reconfigurable gates, are presented in the next section. Besides, the model can be utilised flexibly, where fitting parameters (which have been stated in section \ref{subsec:veriloga} and shown in Figure \ref{Steps}(d)) can be modified to represent other memristor models. In this section, we only access to construct a single proposed memristor with voltage sources as a demonstration of programming and reading the $RS$. 

    \item Simulation setup: Considering that our Verilog-A memristor model calculates $RS$ against time and keeps tracking the change of $RS$, our model is supposed to run in transient simulation to process both write and read operation.
     In this case, we took a single memristor as a simulation example. A chain of pulses was applied across it followed by a triangular wave in Figure \ref{InputSignals} to conduct write and read of memristor respectively. These are the basic setup for calibration and validation analysis in section \ref{subsec:calibration}. The operation steps are as follows:

    \begin{itemize}
        \item The testbench consists of a schematic having one memristor, a single piece-wise linear voltage source (PWL) and a pulse voltage source (Figure \ref{Testbench}). In the schematic, the direction that voltage sources are connected to the `positive' (p) port of the memristor is defined. The positive bias voltage ($V_b>0$) from `p' excites memristor to a higher $RS$, while the $RS$ decreases when positive voltage applies at `n' port. In testbench (Figure \ref{Testbench}), the proposed model is in OFF transition when stimulated with a positive voltage, while a negative voltage results in ON transition. Two voltage sources generate triangular waves and pulses alternately to process that the pulse changes $RS$ of memristor followed by a triangular wave which keeps tracking $RS$. The detail of the input signal has been shown in Figure \ref{InputSignals}. In this measurement for the proposed model, the identical triangular wave is defined as $0.5V$ peak amplitude ($V_{read}$) within $1ms$ duration. The pulse mainly has three variables: width ($t_{w,\Delta R}$), amplitude ($V_b$) and the number of pulses. The detailed simulation and evaluation of these effects on memristor will be given in Section 3.3 with classification.
            \begin{figure}[ht!]
            \centering
            \includegraphics [width=6cm]{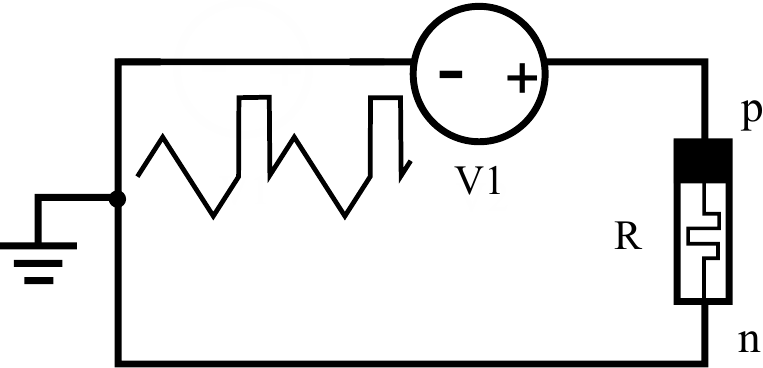}
            \caption{Testbench of applying pulse and triangular wave to memristor to conduct write and read operations in transient simulation.}
            \label{Testbench}
            \end{figure} 
            
            \begin{figure}[ht!]
            \centering
            \includegraphics [width=11cm]{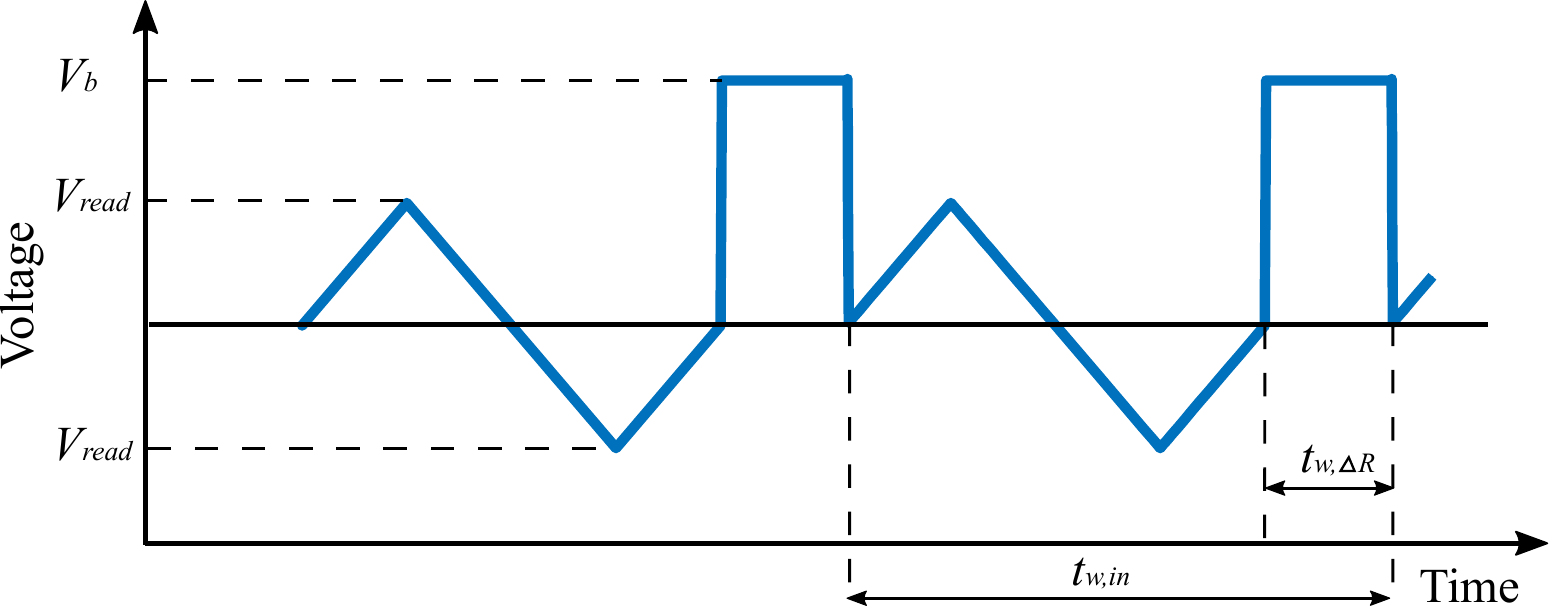}
            \caption{Pulse chain that programme the memristor. The input signal sequence can be divided into two parts: triangular wave and pulse. For all the simulation of our device, the read voltage is defined as $V_{read} = 0.5V$ with $1ms$ duration. The pulses can be determined with specific duration/width ($t_{w,\Delta R}$), amplitude ($V_b$) and numbers of pulses to provoke the memristor. With the combination of two stages, the $RS$ of  memristor model can be tracked for each stimulus.}
            \label{InputSignals}
            \end{figure} 
            \item After setting up the testbench and running the transient simulation, the change of $RS$ is recorded. In this case, each current across the memristor at $V_{read} = 0.5V$ is recorded. With both read voltage and current, the $RS$ can be obtained through division. For detailed operation, users can 1) process the division in Cadence calculator and plot $RS$; 2) send the $RS$ to table and save it as `.csv' file; 3) post-processing the data with Matlab tool and plotting $RS$.
    \end{itemize}    
\end{itemize}

\subsection{Calibration and validation}\label{subsec:calibration}
After introducing the design concept and approaches to integrate the Verilog-A memristor model into Cadence, the model is validated to check whether the Verilog-A model performs in the same way as the physical one. The root mean square (RMS) error for different stimuli is used to validate this. Then, we focused on the performance under specific stimulus in this subsection. 
Combining these validation and calibration parts, this subsection aims at providing users with the operational range and performance of the proposed model, as well as how device parameters affect the above performance. 

Based on the simulation setup and input signal generation in 3.2.2, the performance of Verilog-A model will be validated with physical behaviour extraction in section 2, proving that the proposed model can represent the device finely. Then, we program the proposed model by modulating number of pulse (in 3.3.2), pulse width (in 3.3.3) and amplitude (in 3.3.4) in order to explore both the qualitative and quantitative impacts on model. To be consistent, all simulations will start from the same initial $RS$ as baseline. Recommendation of programming the memristor will be given after evaluating above modulations. 

\subsubsection{Validation}
The Simulation data extracted from our device is presented in Figure \ref{fig:mod-dep} that contains responses based on pulses with different widths for constant and increasing/decreasing amplitude pulses. It is sufficient to prove that the Verilog-A model can be simulated correctly, while the detail discussion based on three modulation pulses will be in the next three sub-sections.\par
As shown in Listing 1, the $RS$ is relative to timestep ($dt$) during simulation. In order to fit the Verilog-A model with physical behaviour finely, the timestep in device extraction and Spectre simulation is supposed to be same, thus, we set the simulation timestep to $1\mu s$. 
%

\begin{figure}[ht!]
\centering
\includegraphics [width=14cm]{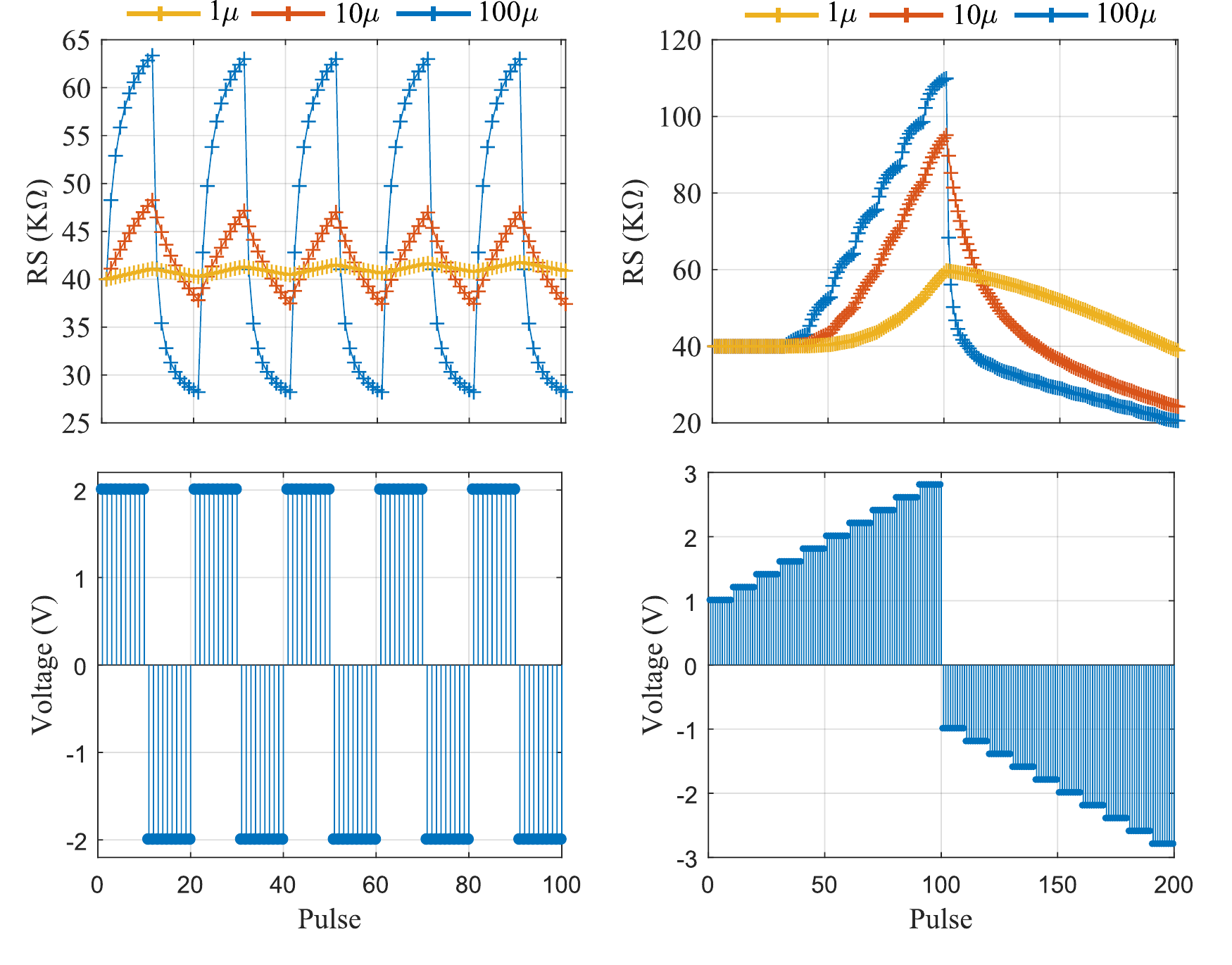}
\caption{Simulation results of Verilog-A memristor model in Cadence. For validation, the same input signals with the one in Figure \ref{fig:mod-dep} with three pulse widths is applied. }
\label{Validation}
\end{figure}

\subsubsection{Number of pulse modulation}
In this part, positive bias voltages: $1.5V$, $1.8V$ and $2.0V$ are applied to the Verilog-A memristor model. These bias voltages are applied to explore the behaviour when memristor is integrated into a CMOS circuit with a nominal supply voltage of $1.8V$. In the meanwhile, $2V$ bias voltage is applied as a typical programming voltage. In our model the boundary of upper limit is depended on the bias voltage for the same initial $RS$ (refer to Listing 1, lines 8, 9 and 40). Thus, we can explore the phenomenon when we applied sufficient number of pulse on the model. Thus, the simulation will be setup as applying three level of amplitude voltages on the device to see the changing of $RSs$ when being provoked by continuous pulses till they are saturated.\par

\begin{figure}[ht!]
\centering
\includegraphics [width=10cm]{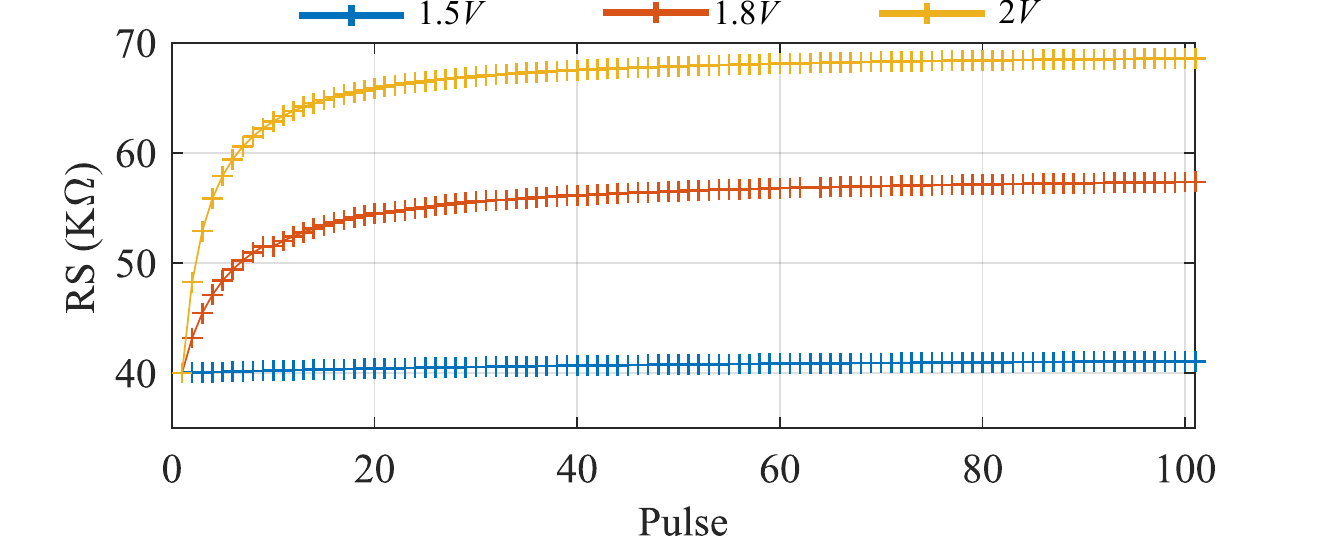}
\caption{Verilog-A memristor model response based on the number of applied pulses. The device is provoked by 100 pulses at different voltage amplitudes - starting from initial $RS=40k\Omega$, and it eventually saturates at different $RSs$ depending on applied voltages. As the pulses keep provoking, changing rate of $RS$ slows down and it gradually saturated. Characterisation routine parameters based on the stimulus in Figure \ref{InputSignals}: $t_{w,\Delta R}=100\mu s$, $t_{w,iv}=1.1ms$, $V_b=1.5/1.8/2.0V$, and $V_{read}=0.5V$}
\label{DiffPulse}
\end{figure}

It can be seen from Figure \ref{DiffPulse} that the rate of change of $RS$ reduces gradually when $RS$ is eventually saturated. At the beginning of the transient, $RS$ changes rapidly and the desired state might be programmed with less accuracy. While a specific $RS$ can be achieved by increasing the number of pulses that helps to program the model in an accurate resistive state. As we are exploring the number of pulse modulation, the effect from amplitude will be omitted temperately. The only discussion relative to amplitude in this subsection is to prove that a sufficient number of pulses can help discover the highest $RS$ for specific bias voltage.\par
\begin{itemize}
    \item By applying sufficient pulses to the Verilog-A model, it can provide users with access to measure the boundary of $RS$ for a specific bias voltage. It helps users to evaluate whether the device can be applied under specific requirements and also helps to define operational voltage on the schematic.
    \item Within $RS$ boundary, a sufficient number of the pulse of different bias voltage can achieve the same desired $RS$. For instance, $RS$ climbs from $49k\Omega$ to $53k\Omega$ by a pulse of $2V$, thus, users need to use $1.8V$ pulses to obtain $50k\Omega$. This induces a trade-off between amplitude and time.
    \item The number of pulses needs to be considered during memristor programming since more number of pulses helps to obtain a specific $RS$.
\end{itemize}

\subsubsection{Pulse width modulation}
\begin{figure}[ht!]
\centering
\includegraphics [width=9.5cm]{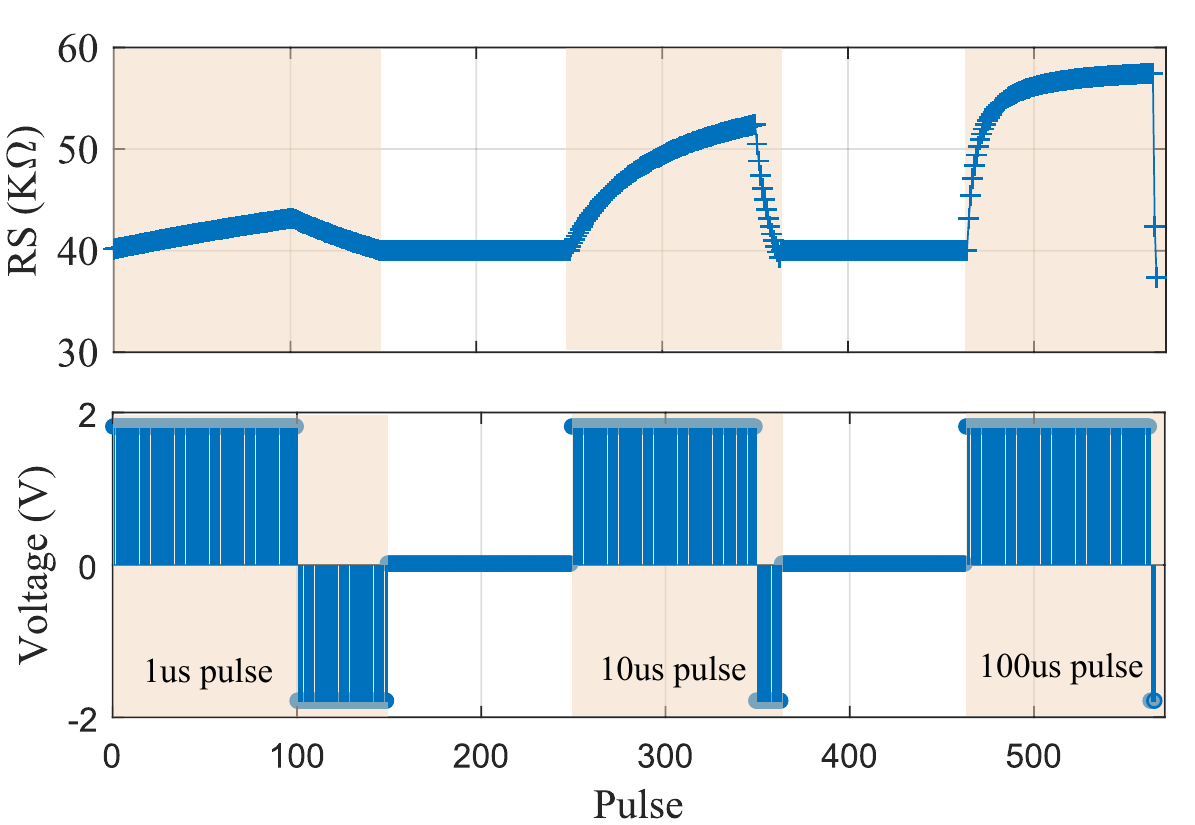}
\caption{Programming the device with three types of duration pulses. Bottom trace:  100 pulses at $1.8V$ with three different duration are employed to modulate device $RS$. In between the measurement, the inverse voltages are applied to flush the device to initial $RS$. Top trace shows the modulation results corresponding to the bottom stimulus with both positive and negative bias voltage. Results with highlight will be shown in Figure \ref{DiffWidth} with a clear view of programming memristor to a specific $RS$. Characterisation routine parameters: $t_{w,\Delta R}=1/10/100\mu s$, $t_{w,iv}=1.001/1.01/1.1ms$, $V_b=|1.8V|$, and $V_{read}=0.5V$.}
\label{DiffWidthTrans}
\end{figure}

Exploration of pulse width effects on programming the memristive devices will be given in this subsection, where we generate pulses at $|1.8V|$, for $1\mu s$, $10\mu s$ and $100\mu s$ pulse width. In Figure \ref{DiffWidthTrans}, three pulse trains with 100 pulses are generated and its equivalent resistive states $RS$ are recorded.  Besides, negative voltage is also employed on the device to flush it back to initial $RS$,  proving that the modulation can be applied in both directions. \par
\begin{figure}[ht!]
\centering
\includegraphics [width=9cm]{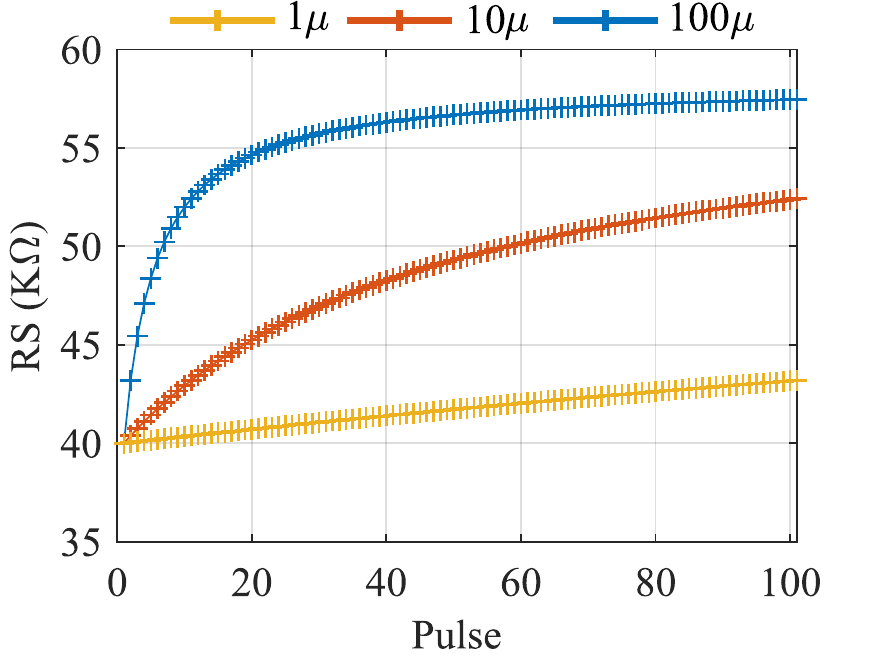}
\caption{Verilog-A memristor model response based on the width of applied pulses. Starting from initial $RS=40k\Omega$, resistive state climbs at different rates for pulses with different widths. The $RS$ increases at a faster rate with more states within first 20 pulses at $100\mu s$ compared with $1\mu s$ pulses. However, the change of rising rate slows down as the number of pulse increases among three situations. It illustrates that the shorter duration pulse helps generate specific $RS$ with higher resolution.}
\label{DiffWidth}
\end{figure}

Figure \ref{DiffWidth} presents pulse width modulation of positive voltage. The $RS$ programmed by $100\mu s$ pulse is more stable and steeper compared with other two pulse width stimulus.
The figure \ref{DiffWidth} indicates that the pulse with smaller width can be applied to slow down the changing rate, which contributes to programming the device to a specific resistive state in application.

\subsubsection{Amplitude and polarity modulation}
\begin{figure}[ht!]
\centering
\includegraphics [width=10cm]{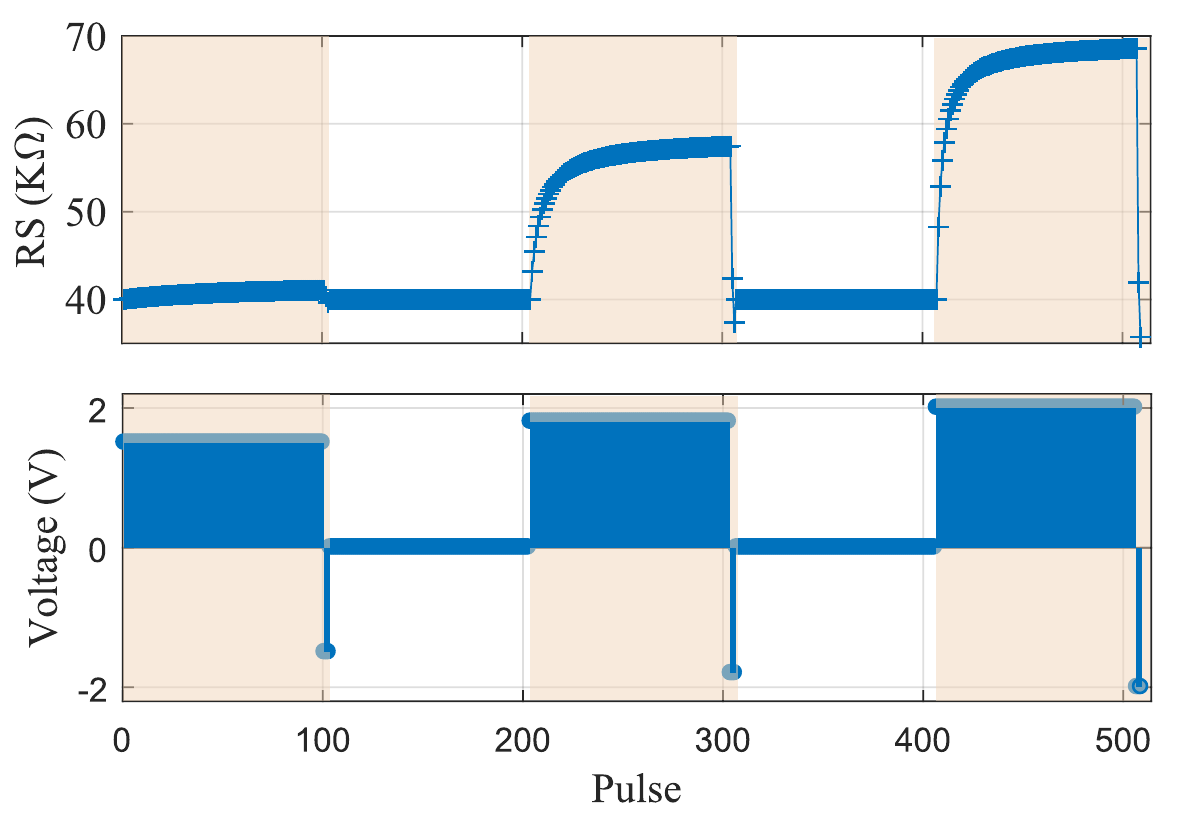}
\caption{Programming the device with three amplitude pulses. Bottom trace: Each  pulse (duration=$100\mu s$) under incremental bias voltage is employed to modulate device $RS$. In between the measurement, inverse voltages help flush the device back to initial state. The $RS$ modulated by positive bias voltage has been highlighted and will be shown in Figure \ref{DiffAmp} for comparison. Characterisation routine parameters: $t_{w,\Delta R}=100\mu s$, $t_{w,iv}=1.1ms$, $V_b=|1.5/1.8/2.0V|$, and $V_{read}=0.5V$.}
\label{DiffAmpTrans}
\end{figure}

To evaluate the amplitude and polarity modulation, a stimulus containing bias voltages at $|1.5V|$, $|1.8V|$ and $|2.0V|$ in two polarities is generated. A 100 pulse trains were applied on the device in order to compare it with Figure \ref{DiffWidth} with fixed pulse number. The negative pluses are applied to pushes the device back to initial state (see Figure \ref{DiffAmpTrans}), which achieves the same $RS$ change with less pulses.

\begin{figure}[ht!]
\centering
\includegraphics [width=12cm]{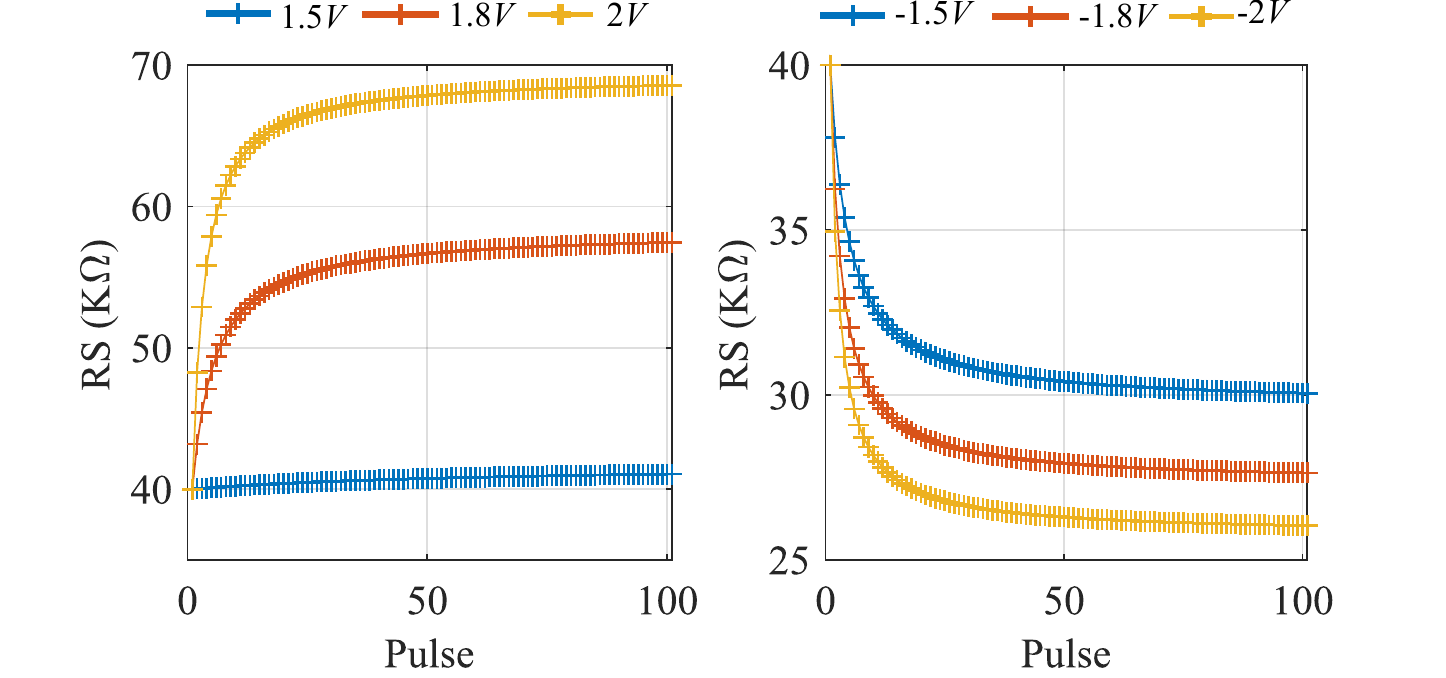}
\caption{Verilog-A memristor model response based on the amplitude of applied pulses. Being stimulated by pulses of different voltage amplitude, simulation results from both positive and negative bias voltages present in left and right figure respectively. It can be seen that the higher absolute voltage leads to faster changing rate of $RS$. Combining two sub-figures, positive voltage induces lower changing rate of $RS$, compared with negative voltage. The right figure also indicates that the saturated $RS$ (boundary of lower limit) is dependent on the bias voltage, where a more negative voltage can push $RS$ to a lower state.}
\label{DiffAmp}
\end{figure}

The higher absolute bias voltage induces faster changing and less pinpointing on the $RS$. Thus, users can select appropriate low voltage to program the memristor and allocate to specific $RS$ level. From Figure \ref{DiffAmpTrans}, it can be found that to reach the same $RS$ level, the number of negative voltage is required much less than positive one. Besides, in Figure \ref{DiffAmp}, negative voltages induce considerable $RS$ drop where the $RS$ cannot be discerned. It is because the device is in OFF transition under positive voltage, while negative bias voltage that lead to ON transition performs faster. It indicates that negative bias voltage is not suitable to determine a specific $RS$ due to the dramatic change. Even though the $RS$ can be dropped to a stable level by applying sufficient negative pulses, it makes effort to measure the specific voltage can achieve desired $RS$.  

Combining above simulation results, some recommendations for application haven been concluded:
\begin{itemize}
    \item Users can apply specific bias voltages with sufficient pulses to explore the operational $RS$ range in simulator at the beginning. The resulting data containing the range of bias voltage, boundary of $RS$ and provoking time, provides users with evidence on whether it can be utilised on their specific schematic and how to apply it.
    \item To program the $RS$ to higher level, positive bias voltage can be applied directly. But there might be two approaches to program the memristor to a lower $RS$: 1) Programming the memristor with negative voltage directly; This decreases the resistance drastically and a desired lower limit is achieved for a specific negative bias voltage. The negative voltage performs in a high speed that $RS$ cannot be recorded.
    2)Dropping the $RS$ to a stable boundary by negative voltage, following by employing positive voltage to allocate $RS$. This is a more practical approach with multiple $RS$ levels being achieved with higher resolution.
    \item To program the device with high resolution, it is recommended to employ a positive bias voltage with a low amplitude and a small pulse width to provoke the device at a slow pace.
\end{itemize}

\section{Designing with Memristors}

\label{Designing with memristors}

All of the effort described in previous chapters sets the groundwork for circuit design using memristive devices (RRAM). Myriads of circuits using RRAM have been conceived and implemented thus far with a broad space of possibilities open for the future. In this section we shall show by examples how design with RRAM can be carried out, pointing out any pitfalls or points of particular interest for the design engineer.

\subsection{Primitive cells for controlling memristive devices}

Once a device model has been settled upon, the next logical step is to start developing primitive CMOS-RRAM cells for practical use as building blocks for larger circuits. At this stage a number of operational, implementation and non-ideal parameters have to be considered. Operational parameters include supported voltages and polarities, pulse durations and in general the waveform parameters of all signals involved in operating the RRAM (which may not necessarily be pulses). Implementation parameters include the sizing of the RRAM and associated MOSFETs, the widths of the back-end lines connecting the circuit and the overall layout topology (for example, do we want the cell to be tile-able into an array?). Layout issues are covered specifically in the next section, but these should be considered during the design phase as well. Finally, non-idealities include the series resistance of transistors, line resistances, maximum voltage tolerances and parasitic capacitances.

In this section we will see a couple of examples of fundamental RRAM-based circuit modules; namely the 1T1R structure that represents the combination of a single RRAM element with a single switch (which can be interpreted as an access control device). This very same structure is used to build crossbar arrays with selectors. Next, we move on to a 2T1R topology used to handle high-voltage electroforming in a CMOS technology that does not easily support high voltages on-chip.

\subsubsection{The 1T1R primitive:} 
\begin{figure}[h!]
\centering
\includegraphics [width=7cm]{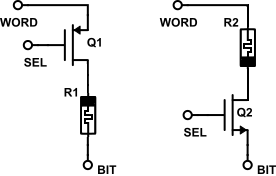}
\caption{Examples of 1T1R topologies. Left: pMOS-based design. Right: nMOS-based design. Note: at this stage we have said nothing about the operation of the 1T1R, so the labels `WORD' and `BIT' are there only as visual guides. Similarly the polarity of the RRAM device is only indicative; in principle the RRAM devices can be arranged or indeed interpreted as being in any direction. The same applies to all schematics in this section.}
\label{1T1Rs}
\end{figure}

We begin with a simple 1T1R topology whereby a transistor is connected in series with a RRAM cell as shown in fig. \ref{1T1Rs}. Despite this being a fundamental building block by itself we shall discuss it within the more complex operating environment of a crossbar array for completeness; this will be reflected in the labelling of the block's terminals, which will correspond to their typical connectivity within the crossbar context. The first issue that arises immediately is whether a pMOS or nMOS transistor should be used as the switching element. This will depend on the importance of the differentiating characteristics of each device respective to the design at hand. nMOS transistors have higher mobility and so can be made smaller for the same current drive capability. Additionally they do not require an N-well substrate structure as a pMOS design would need both N-well taps as well as substrate taps in order to ensure good quality bulk biasing. This makes nMOS devices faster overall, however it also precludes the possibility of using the bulk terminal for anything other than simple grounding. In most cases this will suffice to render nMOS the polarity of choice.

In our example design we have chosen a pMOS-based design for experimental purposes: we wish to be able to control the bulk voltage of each column in the array in order to understand how it can influence device operation. We shall seek to connect the N-wells of the pMOS devices column-wise in our array (which will have implications for layout and so cell size).

Next, we must consider whether we require bipolar operation and if so, whether this should be symmetrical (i.e. we want to pass similar voltage/current magnitudes in both directions). In either case, the required operation voltages/currents will determine the type of transistor that we use; most notably the allowable $V_{GS}$ and $V_{DS}$ (although each MOSFET's $V_{GS}$, $V_{GB}$, $V_{DS}$, etc. all need to be checked thoroughly for suitability). These voltage requirements allow us to exclude any unsuitable transistor types. For example if experimentally a RRAM cell requires 2V to switch, using a MOSFET that will suffer gate dielectric breakdown if either $V_{GS}$ or $V_{GD}$ exceeds 1V would be ill advised. Once a list of suitable transistors is found, the selection can proceed on the basis of `positive attributes', i.e. finding the smallest/easiest to operate/most robust option, depending on project requirements.

\begin{figure}[h!]
\centering
\includegraphics [width=10cm]{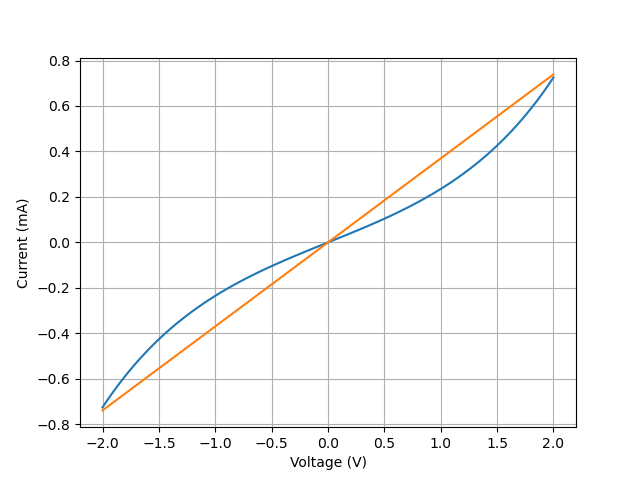}
\caption{Example of a double-exponential IV representing a typical IV behaviour of RRAM devices (blue) together with worst-case linearised IV (orange) within the voltage operating range of interest; here [-2,+2]V. By using the orange IV we always assume the device will require more current than it actually does at any given bias voltage.}
\label{IVexams}
\end{figure}

In our example we will consider the symmetrical case. The RRAM devices will be modelled as having a minimum resistance of $1\,k\Omega$, fixed for all voltage biases across the device, and require $1.5\,mA$ to operate in both directions; in other words, we always consider the device to be at its absolute worst case (see toy example in Fig. \ref{IVexams})\footnote{This is a very conservative design approach and the level to which it is secure depends on the confidence level that the actual worst case is known. In a technology characterised by uncertainty, determining the confidence interval is an activity that must be undertaken very carefully. Furthermore, linearising the IV to worst case is useful for setting design constraints. It should not be used for precise design as the estimated and actual currents may vary quite significantly.}. This translates to 1.5V dropped across the worst-case memristor in both directions. This means that the transistors used in the 1T1R scheme will have to be rated for at least 1.5V. In practice the required rating may become substantially higher if the series resistance of the FET device is comparable to that of the worst-case RRAM resistance. Note: we can always trade off transistor aspect ratio for FET series resistance, but there are practical limits to how wide we are willing to design our transistors. In our case this means we limit ourselves to using 3.3V or 5V devices; excluding all lower voltage variants offered by the technology (for the required transistor width becomes exceedingly large if our power supply headroom is reduced to, say, 2V).

\begin{figure}[h!]
\centering
\includegraphics [width=7cm]{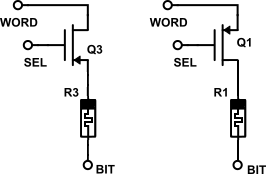}
\caption{Circuit diagrams for `source-to-RRAM' (left) and `drain-to-RRAM' (right) connectivity configurations. These are not identical if the transistor is not symmetrical (i.e. its drain and source terminals are not interchangeable).}
\label{TRRT}
\end{figure}

Once a suitable transistor has been selected we need to identify whether is it symmetrical or not, i.e. whether source and drain can be used interchangeably. Should that be the case, the 1T1R design becomes symmetric up to relabelling. However, should that not be the case (e.g. in the case of an extended drain transistor), the $V_{GD}$ required to admit some drain current $i_x$ will differ from the corresponding $V_{GS}$. In this case the natural first instinct would be to `point the source' of the transistor towards the RRAM as shown in fig. \ref{TRRT} (left schematic) so that when current flows first through the FET device and then through the RRAM, the FET drops approximately its $V_{GS}$' worth of bias voltage, leaving the rest to drop across the RRAM (saturation mode assumed). Then, when the polarity of the current is reversed, $V_{GD}$ (now acting as the effective $V_{GS}$) is directly controlled by $V_{G}$ and one of the biased lines (wordline or bitline), i.e. it no longer depends on the voltage between FET and RRAM. This allows the FET to now consume $V_{DS} \geq (V_{GD}-V{th,D})$ bias voltage while remaining in saturation, where $V_{th,D}$ is the effective threshold voltage when the FET is used `in the wrong direction', with $V_{GD}$ acting as $V_{GS}$. Generally $V_{th,D}$ can be expected to be substantially larger than the `normal' $V_{th}$. As a result, the `point the source towards the RRAM' topology seems to allow for a more balanced application of voltage/current across the RRAM device all else being equal by mitigating the effect of the elevated $V_{th,D}$ \footnote{Of course, in both cases significant channel resistance (the $V_{DS}$ effect) will add a further layer of complications to this design.}. This analysis assumes that the effective voltage threshold is lower when the devices is operated with the source and drain terminals playing the intended roles, rather than being reversed - i.e. that the devices are asymmetric. Table \ref{TRRTtab} illustrates this situation by showing numbers from a simulation of both configurations' behaviour under 5V power supply and 1$k\Omega$ fixed resistive load. It is clear that the `source-to-RRAM' configuration is more balanced and thus more suitable for RRAM devices with symmetric current requirements, such as we investigate in this example.

\begin{table}[!h]
\centering
\caption{Current passed through 1T1R configurations for the `source-to-RRAM' and `drain-to-RRAM' circuit configurations for an asymmetric pMOS transistor feeding a 1$k\Omega$ load under a 5V supply. Currents are given in the forward (current flows from `WORD' to `BIT' terminals) and reverse voltage bias regimes. It is clear that `source-to-RRAM' supports that highest minimum current, but `drain-to-RRAM' supports the highest maximum current. Asymmetric devices were used for this test.}
\begin{tabular}{l|c|c}
               & Forward & Reverse \\ 
\hline
Source-to-RRAM & 2.5mA   &  1.7mA    \\
Drain-to-RRAM  & 4.5mA   &  1.3mA  
\end{tabular}
\label{TRRTtab}
\end{table}

However, the drain-to-RRAM configuration could be useful for cases where the RRAM devices have highly asymmetric current requirements. The reason is that the drain-to-RRAM configuration supports substantially higher maximum currents in the forward bias case (i.e. where current flows from `WORD' to `BIT' terminals). Thus, the conclusion is that unless the situation has a very obvious solution (e.g. there are obvious symmetries or extremely accentuated asymmetries), it is advisable that simulations are carried out for both topologies and both bias polarities before the most acceptable 1T1R configuration can be selected.

The simulations used to generate the corner cases summarised in Table \ref{TRRTtab} are very simple: They are DC operating point analyses of the circuits illustrated in fig. \ref{TRRT} taken at the corner cases where the voltage across the 1T1R stack is the nominal power supply (typ. close to the maximum voltage tolerated by the transistor) and the gates of the transistors are fully open ($V_{G}$ = VDD for nMOS devices and GND for pMOS\footnote{Note: in the more general case where the drain terminal of the device can exceed the power supply rails the formula is for $V_{gs}$ = VDD, so that the allowable values of $V_g$ depend on the current value of $V_s$}). A specific configuration and transistor sizing will in general pass the basic performance criteria if: a) It can successfully pass sufficient current in both directions onto the worst-case load and b) It falls within the rated voltages of the devices supported by the technology. Note: in some technologies this is indicated by the presence or absence of safe operating area check (SOAC) errors.

Thus far we have covered the signal voltage and polarity support. The pulse duration then introduces an additional set of considerations, mostly in terms of heating. Continuous stress, as might arise due to extremely long pulses (possibly even ms or s) means that the calculations for FET, and even more pertinently metal line current-carrying capacity, need to be adjusted accordingly. In general, however, this becomes a layout issue for the back-end lines as transistor sizes are adjusted to keep current densities (and as a result channel resistance and local power dissipation) under check.

Finally, we note that just like in any other array design, aiming for good layout compactness, matching, low parasitic capacitance and tileability farther down the design cycle helps improve the performance of the 1T1R cell. This is particularly pertinent in 1T1R designs that are very frequently used as the building blocks for large crossbar arrays.

Please note that frequently in practical designs, the assumption of symmetrical FETs can be made, which greatly simplifies the design.

\textbf{Additional, crossbar array-specific points:} As alluded at the beginning of this section, if the 1T1R primitive is to be used in a crossbar array, additional design considerations arise, most notably the interconnection of the 1T1R elements. Whilst this lies outside the scope of this section, we will note that the typical connectivity of a crossbar stipulates (canonically) that the wordline (connecting all `WORD' terminals) runs perpendicular to the bitlines and selector lines (shorting BIT and SEL terminals column-wise respectively). This allows us to use combinations of (WORD, BIT) and (WORD, SEL) to isolate individual RRAM devices within the array with the aid of their corresponding `selector transistor'. Other approaches are possible, but in general maximum versatility is ensured if one set of lines runs perpendicular to the other two.

Additionally, the connectivity of the transistor bulks needs to be specified. In a vanilla nMOS-based array, the bulk terminal will be the substrate and so there is no further decision point. However, if pMOS devices, triple-well nMOS or e.g. FDSOI technologies are used the connectivity pattern of the bulk terminals needs to be also specified. In general, there are only a handful principal connectivity combinations that satisfy the restriction that for an $N \times N$ array we have a maximum of N bulk lines servicing all the bulks. First, column-wise connected bulks would allow crosspoint combinations between bulk and wordline bias in a similar fashion as is used to isolate individual cross points for normal word-line/bit-line operation. Next, we could have row-level connectivity. Finally, there is the option of connecting all bulk terminals to a common supply.


\subsubsection{The 2T1R primitive:}

Next, we consider a more complicated primitive cell example. This is an illustration of how primitive cell design can quickly become much more complicated if the RRAM technology used requires voltages higher than what the underlying CMOS technology supports as a standard.

In this scenario we have a technology that supports up to X volts as a standard, but the RRAM requires $Y \gg X$ volts bidirectionally for guaranteeing successful electroforming \cite{Stathopoulos_2017}. In our scenario the CMOS technology does not feature transistors that support Y across all possible terminal pairings, but does feature extended drain FETs that support $V_{DS}$, $V_{GD}$ and $V_{DB}$ at the level required by Y. Furthermore, the resistance of the RRAM device is assumed to be able to take any value, including single digit kOhms and MOhms. For the example that follows we arbitrarily set X=3V and Y=10V.

\begin{figure}[h!]
\centering
\includegraphics [width=4.5cm]{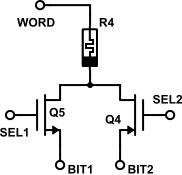}
\caption{Schematic of the 2T1R system with asymmetrical devices, that we use to bidirectionally apply `high voltages' (above the supported max. $V_{GS}$ of the transistors) to the test RRAM device. BIT1 is grounded whilst BIT2 is fixed to the highest voltage we wish to apply (here 10V). When we need to apply `positive' voltage to the device, Q5 is ON and WORD is at voltage $V_w>0$ as appropriate. When applying `negative' voltage to the device, Q4 is ON and WORD is at $V_w<10$ as appropriate. Q4 and Q5 are never concurrently ON.}
\label{2T1R}
\end{figure}

By the assumptions of our scenario we cannot guarantee a sufficiently low voltage drop across the RRAM cell. As a result, a 1T1R solution is unworkable even using the extended drain devices because we would be forced to assume that the transistor's $V_{DS}$ can swing between $\pm10V$. Instead, we opt for a 2T1R approach as shown in fig. \ref{2T1R}, with extended drain (and therefore asymmetrical) devices. The circuit operates on the basis that the two bitlines are fixed at 0V and 10V while the wordline can move freely between 0-10V. This might be the case if the wordline is directly connected to a pad (with appropriate ESD voltage limits), for instance. We note that both FETs have their sources `pointing away' from the RRAM, which is done in order to guarantee that by controlling $V_{G}$ we can control the `normal' $V_{GS}$ in both cases (since $V_{GS}$ must be $<$3V). Then, when we wish to apply +10V to the RRAM, we apply 10V on the wordline and sink the current through the nMOS Q5. Similarly, for applying -10V on the RRAM, we ground the wordline and turn on the pMOS (Q4). For intermediate voltages we set the wordline voltage correspondingly and choose according to the desired polarity which transistor will be responsible for applying the voltage. Naturally, how much of the voltage drop actually does occur across the target device is something that needs to be checked thoroughly and simulated.

In the example above, the pMOS and nMOS are never operated simultaneously. Furthermore, the gate voltages swing between [0,3]V for the nMOS and [7,10]V for the pMOS, therefore appropriate circuitry guaranteeing that no $V_{G}$ can exceed those limits needs to be designed. Whilst this is no longer part of the primitive design, it is an example of a key primitive core decision that significantly impacts the design of the rest of the system. Transistor selection and other implementation and parasitic considerations are similar to the 1T1R case with appropriate modifications and are left for the reader.

Finally, we note that in this example we chose to keep both bitlines at fixed voltages (GND and 10V), thus naturally extending the 1T1R case where the bitline is fixed to GND. There may, however, be interesting control options in varying those voltages as well.

\subsubsection{Other primitives:}

\begin{figure}[h!]
\centering
\includegraphics [width=5cm]{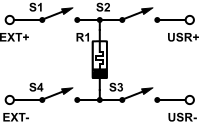}
\caption{Schematic of double transmission-gated RRAM primitive cell. In normal operation either the pair S1,4 or S2,3 are closed, connecting the RRAM cell to the `external' (EXT terminals) or `in-circuit' environment (USR terminals). This is very convenient in particular for circuits that only require the RRAM devices to maintain a fixed resistive state during circuit operation.}
\label{TGRRAM}
\end{figure}

Naturally, there are other significant RRAM-CMOS topologies that can be developed, such as a dual transmission gate-accessible RRAM cell (fig. \ref{TGRRAM}) that can be used to switch the RRAM cell between `in-circuit' and `out-of-circuit' operation. This could be used for example to operate the RRAM cell normally in-circuit and then change its resistive state in a very controlled fashion using the out-of-circuit terminals.

\subsection{Simple design example: a reconfigurable gate}

\begin{figure}[h!]
\centering
\includegraphics [width=15cm]{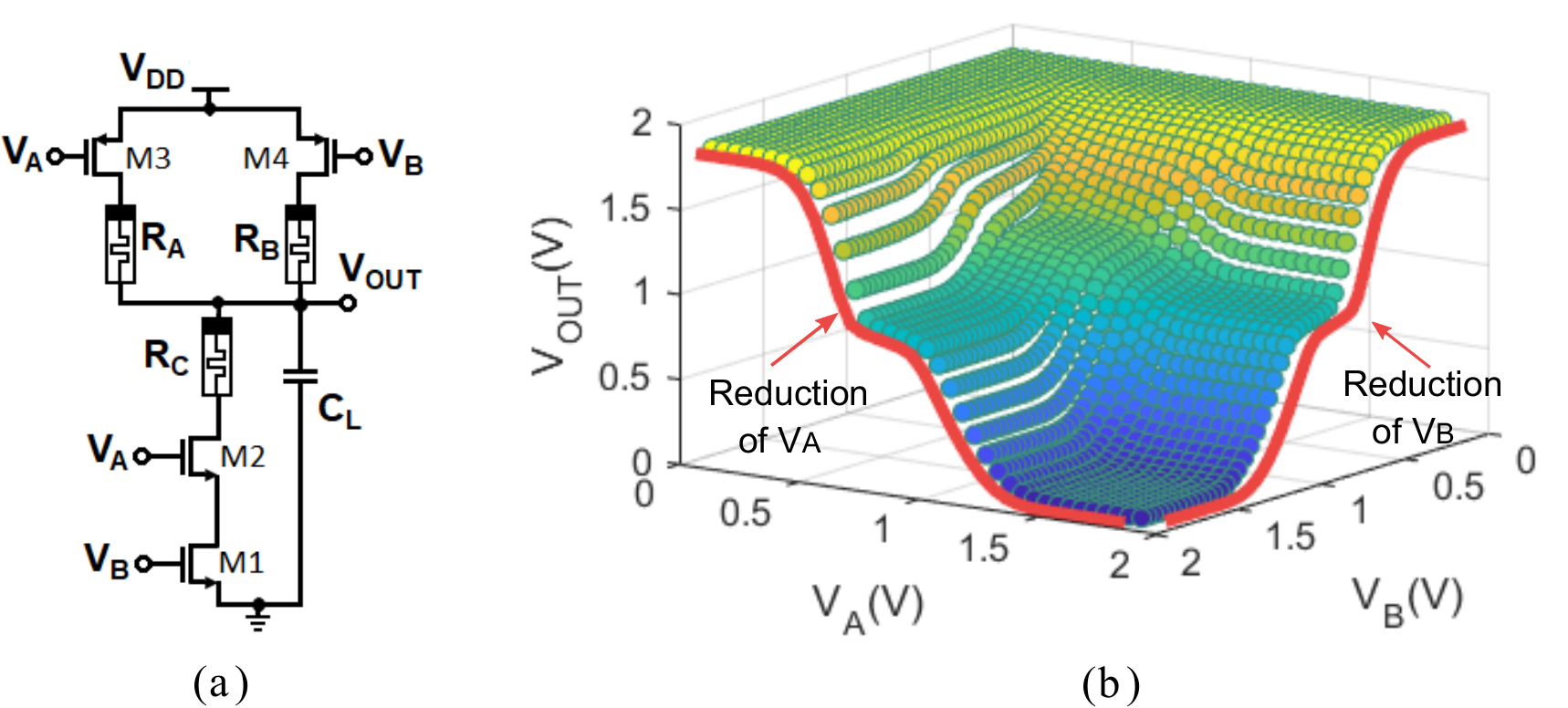}
\caption{RRAM-enhanced NAND gate. (a): Schematic of the RRAM-enhanced gate. (b): Input/output transfer function for a number of (analogue) input voltages at $V_A$ and $V_B$. Changing the values of $R_A$, $R_B$ and $R_C$ changes the shape of the transfer function. Corner simulations (with extreme values of $R_{A,B,C}$) can reveal the extent to which the transfer function surface can be manipulated whilst simulations with $R_{A,B,C}$ at mid-range provide a good indication of the transfer function shape with the RRAM devices at their most flexible. Reproduced from \cite{Serb_2018}.}
\label{NANDR}
\end{figure}

The use of our primitive components is exemplified very clearly in the design of a reconfigurable logic gate, as originally described in \cite{Serb_2018}. The schematic for the reconfigurable NAND gate is shown in fig. \ref{NANDR}. We notice that the NAND gate can be decomposed into a set of 2x pMOS-type 1T1R cells plus a single nMOS-type 1T1R with an extra transistor for enforcing the pull-down path. Naturally the 1T1R structures can be implemented such that the places of the RRAM and the transistors are swapped, as required by the problem specifications.

\subsubsection{Nominal circuit design:} Once the basic architecture of the gate is decided, the circuit needs to be fully specified (e.g. transistor sizes and RRAM nominal resistive state ranges). Importantly, we talk about RRAM `resistive state ranges', as opposed to simply `states' because in the general use case the reconfigurable gate will operate in a `lifelong reconfiguration' mode, as opposed to what one might call a `configure \& forget' modality. Thus, in the general case the design will require that the I/O transfer characteristic can cover a multitude of states within certain bounds, as opposed to simply taking a single nominal shape.

The procedures for elaborating 1T1R-based circuits are still developing, but a good starting point is to set all CMOS devices to a suitably chosen sizing\footnote{This may be such that the nominal circuit with all RRAM devices at the middle of their resistive ranges directly yields the central-case nominal I/O transfer characteristic, or set to minimise overall circuit size, or anything else as determined by the application specifications.} and all RRAM devices to the middle of their expected operating range and then extract a 3D plot of the gate output voltage against the voltages at the inputs (A,B). If the resulting performance is acceptable, the corners can be analysed next: pull-up RRAM devices are set to their maximum (minimum) allowable values and pull-downs to their minimum (maximum) values and the shape of the input/output (I/O) function is examined. Thereafter, an iterative process (either manually, or using machine learning techniques) can be carried out for altering the RRAM resistive state ranges until the I/O function meets all specifications. This may imply running the iterative process once for every `corner' case of the I/O transfer characteristic, where `corner' implies that the chosen I/O characteristic places a unique restriction on the RRAM resistive range of at least 1x RRAM device. The union of all resistive ranges demanded of RRAM device X by all corner cases yields the set of resistive states that device X should cover to guarantee fully correct nominal operation. Should the I/O function specifications require RRAM resistive ranges lying outside the operating ranges supported by the initial choice of RRAM device designs (i.e. a solution cannot be found), either the RRAM devices can be redesigned or transistor sizing can be changed\footnote{Note: In order to develop better intuition of the problem, the following may also be carried out in aid of the designer: For both pMOS and nMOS 1T1R types, plot the current through the 1T1R vs. the voltage applied at the gate of the transistor and the voltage at the output node of the gate. This creates plots that can be interpreted as 3D load-lines.}.

Once the gate is complete, the usual battery of tests used to check circuit functionality and reliability can be applied to it: i) DC sweeps to check the I/O transfer characteristic, ii) transient analysis to check rise/fall times and assess power dissipation, iii) Monte Carlo (MC) and corner analysis to assess the full performance profile, iv) temperature sweeps etc. CAUTION: When carrying out any of the above tests it is imperative that the limitations of the RRAM model are taken into account: whilst the CMOS component of the circuit will account for all of the above behaviours, not all RRAM models will include e.g. MC or temperature data (almost none do at the time of writing).

\subsubsection{Dealing with uncertainty:} Importantly, during MC and corner analysis the RRAM devices are treated slightly differently than ordinary CMOS devices because of the reconfigurability of RRAM. When designing our nominal circuit we explained how the requirements of the set of desired I/O characteristics in general lead to a resistive \emph{range} requirement for each RRAM device. There is no guarantee that the device's actual operating range can cover the desired range, however. Furthermore, in general, a correction factor needs to be included to account for uncertainty in the devices' actual resistive state ranges. Thus, if our circuit analysis has resulted in a desired RRAM range of [A,B] (derived from all sources of uncertainty except the RRAM device itself), the nominal RRAM resistive range is [X,Y] and the $\alpha$ percentiles of the RRAM resistive range extrema lie at $X+q$ and $Y-q$\footnote{The use of `percentile' here is illustrated by the following example: We have a RRAM with nominal minimum resistance $R_{min}=1k\Omega$, but 90\% of devices in the technology will have an actual $R_{min}$ of at most $1.15k\Omega$, then we say that the 90th percentile of $R_{min}$ is $1.15k\Omega$.} we must ensure that:

\begin{equation}\label{uncertaintyEq}
    [A,B] \in [X+q,Y-q]
\end{equation}
in order to guarantee that our RRAM-induced yield limit is at least as high as $\alpha^2$/device.

For this task, RRAM models including a description of (practically reachable) resistive range variability become necessary. As per standard practice, if the specs are not met under uncertainty, the system needs to be adjusted accordingly.

The exact same approach holds when we introduce operating temperature as a parameter in determining corners. This approach can, in principle, be automated and integrated into current industry-standard CAD tools such as Cadence; the principle of operation would be similar to the current criterion-based testing tools available for pure CMOS designs. In that respect designing with RRAM is no different than designing with pure CMOS.

\begin{figure}[h!]
\centering
\includegraphics [width=7cm]{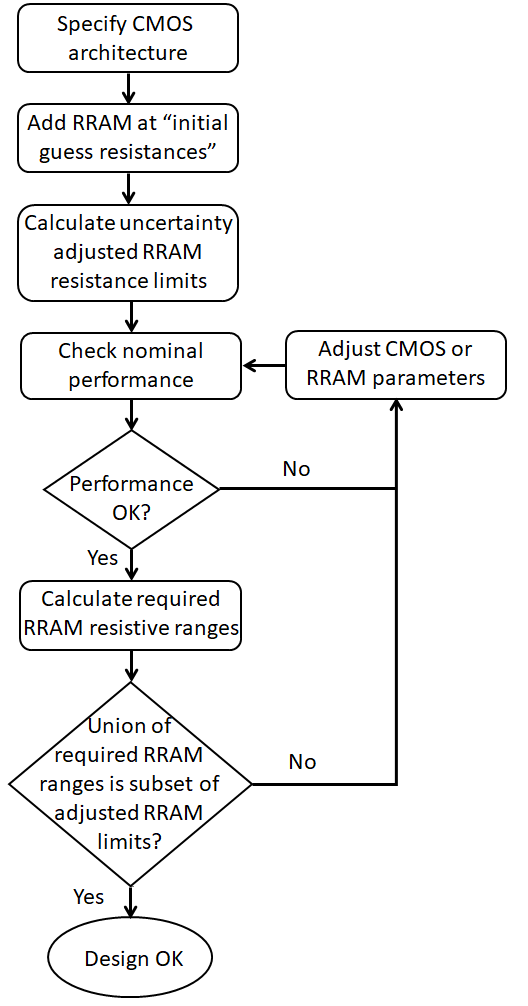}
\caption{CMOS design with RRAM indicative workflow. The preparatory work entails specification of the overall circuit architecture, adding RRAM devices at some nominal, initial guess states and computing the uncertainty-adjusted limits of the RRAM technology in use. Then the iterative part of the design begins. First, nominal performance is checked and adjustments made if necessary. Then, the required ranges for each individual RRAM device are computed such that the circuit can cover all its performance  corners. If the union of required RRAM ranges is a subset of the originally calculated uncertainty-adjusted RRAM limits, then the design can proceed. Otherwise, further adjustments need to be made.}
\label{dFlow}
\end{figure}

In summary, a good design flow taking into account the reconfigurability and uncertainty of RRAM can be presented as per the flow chart of fig. \ref{dFlow}.

\section{Physical Design - Layout}

After we have designed our CMOS-RRAM primitive cells and larger systems, this section moves to the physical design of memristor devices and systems, referring to the design flow (fig.\ref{ProcessFlow}) shown in Section 3. The layout approach is in conjunction with the existing CMOS based layout rules. The versatility of the approach is its compatibility with CMOS rule set, thus the memristor layer can be retrofitted with the existing process. The CMOS routines are linked to the memristor layers through customised vias allowing the memristor to be regarded as a standard component in the circuit.

In this section, the layout of a single memristor is firstly introduced and mapped to its physical structure. The environment setups enable the Layout Versus Schematic (LVS) and Parasitic Extraction (PEX) simulations for memristors. Moreover, several layout examples for memristor primitives depicted in Section 4 are demonstrated. A further step into layout design involving memristor array is provided. Finally, this section demonstrates memristor layer mapping description for exporting layout to GDSII stream. The layout examples provided in this section are constructed in 0.18\textmu m CMOS technology and the physical verification is performed with Calibre. 

\subsection{Standard layout cell}
\begin{figure}[ht!]
\centering
\includegraphics[width=15.5cm]{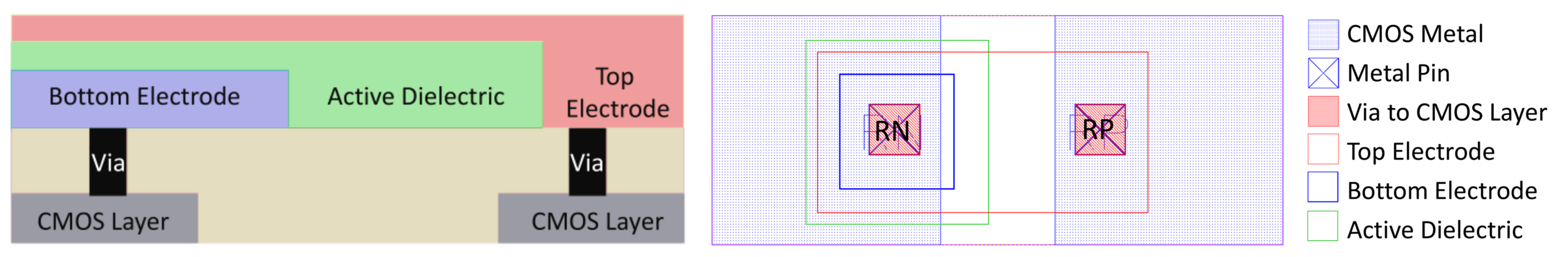}
\caption{Layout of a memristor(right) with its equivalent cross section view (left). Layers are annotated corresponding to its MIM physical structure. Vias to CMOS layer create the link between memristor and CMOS metal routing. Memristor Boundary is a reference layer that is not manufactured.}
\label{layout_rram}
\end{figure}

Memristors, as a new designed devices, require a fully customised layout design by engineers to specify the layers of different functionalities. In order to separate them from the CMOS process, the model should be built on layers that differ from CMOS manufacturing layers. Fig.\ref{layout_rram} shows the layout of a single memristor with its layers labelled corresponding to its physical structure. An equivalent cross section view is provided on the left.

Figure \ref{layout_rram} on the right displays the layout of our in-house memristor implementation, including four memristor specified layers and one metal layer from CMOS. The layer Memristor Boundary delimits the area of the memristor as a reference layer which is not manufactured. This layer has a dimension of 5 \textmu m $\times$ 2 \textmu m. The layout also has a 2.9 \textmu m $\times$ 1.4 \textmu m top electrode and a square bottom electrode with a size of 1\textmu $m^{2}$. Meanwhile, the active dielectric lies in-between both electrodes, covering a slightly larger square than the bottom electrode (1.6 \textmu m $\times$ 1.6 \textmu m). There are two vias between memristor metal layers to higher CMOS metal layer, enabling the access to top and bottom memristor electrodes in CMOS process. Finally, two CMOS metal pins labelled as RP for memristor top electrode and RN for its bottom one are also placed. As a result, the memristor can be linked to CMOS design as a standard cell by routing the wanted memristor port to CMOS metal wires.

The memristor layout in fig \ref{layout_rram} will be used as a standard layout cell in the following examples. In this cell, the CMOS metal layer, metal 4, is used to provide connections between memristor and CMOS routes. However, this layout configuration is not the only way to design a memristor; other variations are also possible. The dimensions of each layers can be adjusted depending on the technology that is used to fabricate the memristors, as well as other factors such as design purposes. At the same time, the distance between the two vias is also configurable. Some other configurations do not use the top electrode layer at all because the device is connected directly to a CMOS pin rather than a customised metal pin. Situations can be highly variable according to different design. 

\subsection{Layout instruction and calibre environment setup}
\subsubsection{Calibre design rule check}
To fabricate the design successfully, DRC files are provided by the foundry that help designers check any design violations. It is the first step before running the Layout versus Schematic (LVS) check. Performing DRC checking regularly can avoid accumulated errors.

\subsubsection{Calibre layout versus schematic}
Calibre LVS compares electrical circuits from the specified source netlist and layout geometry. LVS applications establish a one-to-one mapping between the elements of one circuit (instances, nets, ports, and instance pins) in the source netlist to the layout circuit. This matching is completed when a one-to-one mapping between the elements is established. However, the standard pdk is unable to recognise the memristor device as this is not a built-in device in the design kit. Thus, when the LVS application attempts to map the source netlist to the layout netlist error are reported.

In order to generate an equivalent matched circuit, the description in the text box below is required to be appended into the original runset file located in the design kit. Additionally, the design kit contains another file named \textbf{“source.added”}. This file contains the information of all the sub circuits in the design kit. Here, the declaration of the memristor device used in the design is utmost important as otherwise LVS will show an error.
 


\vspace{1mm}
\begin{Verbatim}[frame=single, label=\large\textrm{{calibre.lvs/calibre.rcx (runset file)}}, framesep=3mm, fontfamily=tt]
LAYER MEMRESLYR          450 
LAYER MAP 215 DATATYPE 21 450 //  layer to form memresistor
MEMRESLYRT = MEMRESLYR AND M4
MEMRESLYRZ = MEMRESLYR NOT M4
CONNECT metal4 MEMRESLYRT
DEVICE memristor MEMRESLYRZ MEMRESLYRT(RN) MEMRESLYRT(RP) netlist model memristor
\end{Verbatim}

\vspace{1mm}
\begin{Verbatim}[frame=single, label=\large\textrm{{source.added \textbf{XXXX}}}, framesep=3mm, fontfamily=tt]
.SUBCKT memristor RN RP
.ENDS
***************************************
\end{Verbatim}

\subsubsection{Calibre netlist extraction}
After successfully running DRC and LVS, Calibre xRC (PEX) is required to generate extracted view with all the parasitics. 
For creating PEX (xRC) rule files, same regulations (as for “calibre.lvs” calibre rule runset file) should also be appended to the “calibre.rcx” file. Moreover, in order to extract the memristor netlist in the calibre extracted view, another description needs to be written and appended in the original “calview.cellmap” file present in the design kit. 
NOTE: The presented model does not incorporate the parasitic model hence PEX will only extract the netlist for the memristor without any parasitic. 


\vspace{1mm}
\begin{Verbatim}[frame=single, label=\large\textrm{{calview.cellmap}}, framesep=3mm, fontfamily=tt]
(memristor
  (std_memristors memristor symbol)
  (
    (RN RN)
    (RP RP)
  )
  (
    (nil multi 1)
    (nil m 1)
  )
)
\end{Verbatim}

\subsection{Layout of primitive cells - 1T1R and 2T1R} \label{1t1r2t1r}
Once we have designed a standard cell for memristor, systems involve memristors can then be normally routed as they are in CMOS process. The standard memristor layout cell used in this paper only has one mask for CMOS fabrication, metal 4 layer. Any routing on this layer should be carefully handled to avoid unwanted shorts. Normal Layout routing tips for CMOS fabrication still apply to the systems here, for example, minimising noise coupling on critical signals, matching differential signals and widening power rail.

Layout of memristor system varies as this is dependent on the circuits being designed, and the layout of the cell. To illustrate the differences, layout examples are given based on the designed 1T1R and 2T1R primitives discussed in Section 4. These systems only include a single memristor whose layout design can be simple as long as only one memristor is used in a design. A more complicated case, the crossbar array, is discussed later.

\begin{figure}[ht!]
\centering
\includegraphics[width=15cm]{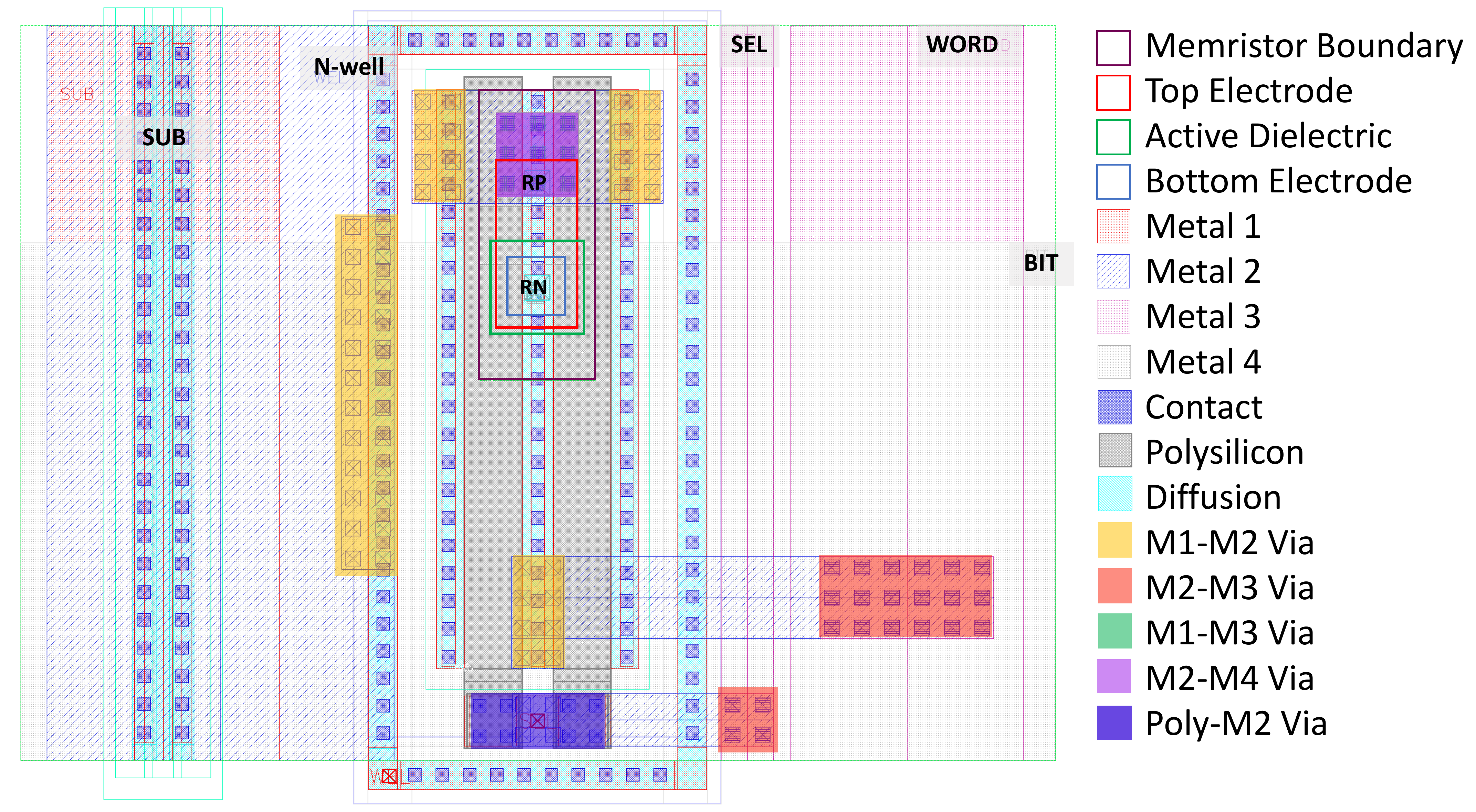}
\caption{The 1T1R primitive Layout with a focus on the memristor-linked routing. Its schematic is provided in fig.\ref{TRRT} (left). The net names corresponding to the schematic are labelled on the layout. The memristor is placed above the pMOS with metal via M2-M4 (3 $\times$ 3) connected to its top electrode. The vias are then routed by metal 2 and via M1-M2 (4 $\times$ 2) providing necessary joints to transistor source terminals. Body connection is placed next to the pMOS. This graph only displays mostly the routing layers.}
\label{layout_1t1r}
\end{figure} 

Fig.\ref{layout_1t1r} shows the 1T1R layout structure in detail. Its corresponding schematic is given in fig.\ref{TRRT} (left) in section 4. The three nets (BIT, WORD, SEL) corresponding to the schematics are labelled on the layout, as well as the pMOS body. The layers of memristor are provided for references. To make the design more visible, only selected CMOS layers are displayed, mostly the routing layers. In the 1T1R structure, the layout is made up by only a pMOS and a memristor. The pMOS has a much wider width than the memristor, so it is split into two fingers with a shared drain terminal. The memristor is placed above the split transistor that shares the drain terminal. A zoomed insight to the memristor-related connections is provided in the blue box in fig.\ref{layout_1t1r}. According to the schematic, the memristor is connected to the pMOS source terminal. This connection is achieved by metal via M2-M4 (3 $\times$ 3) on the memristor RP port and metal via M1-M2 (4 $\times$ 2) on both sides of the pMOS source terminals. The via-to-via route (M2-M4 via to M1-M2 via) are connected by metal 2 layer. Moreover, the bottom electrode RN of this memristor remains its connection to the only CMOS layer that in the memristor standard cell, metal 4 layer. This electrode further has a metal 4 pin labelled C matching with its port on schematic. Additionally, this pMOS-based 1T1R design has greater area occupied compared to the same size nMOS-based 1T1R cell due to the extra body connection for pMOS. 

\begin{figure}[ht!]
\centering
\includegraphics[width=15.6cm]{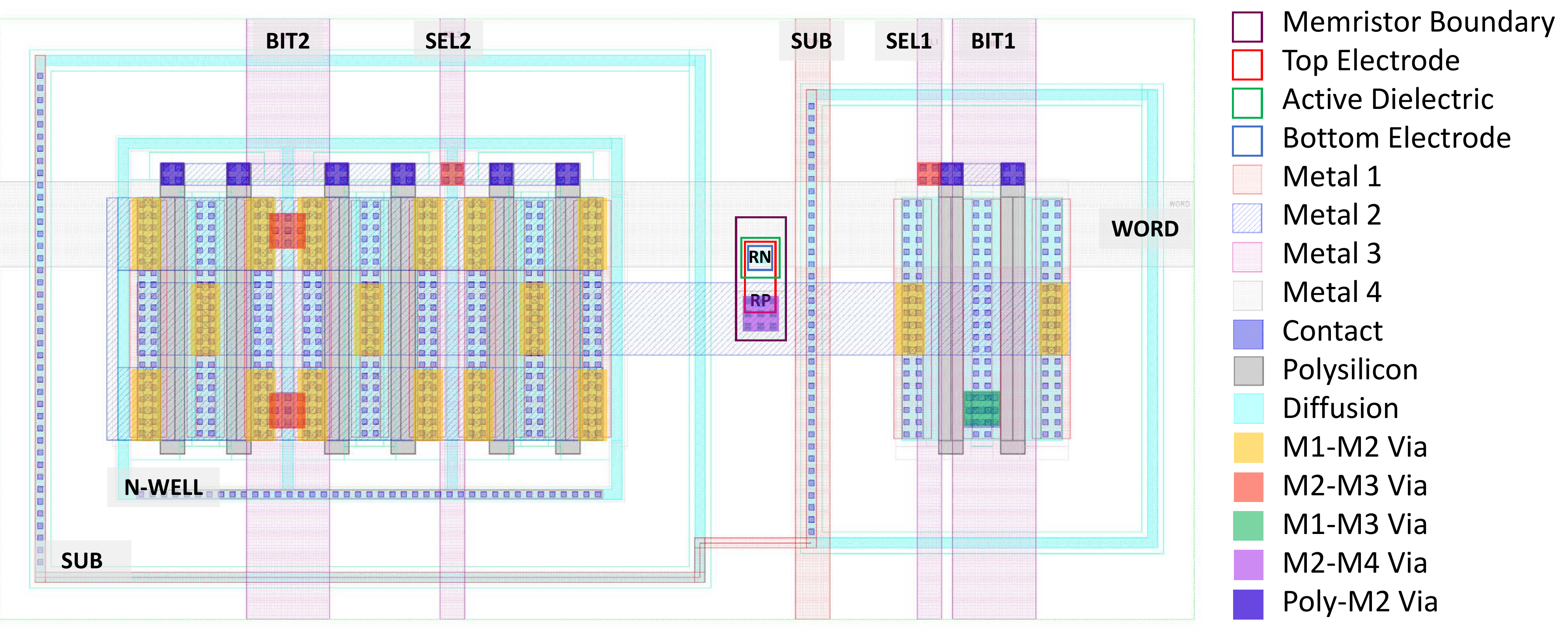}
\caption{The 2T1R primitive Layout highlighted with M1-M2 via on transistors. The schematic is given in fig.\ref{2T1R} with the same net names labelled. A zoomed view of memristor area showing metal via M2-M4 on the top electrode. All the signals for ports are placed till the edge of the cell bringing benefits for array routing. This graph only displays selected layers for a clearer view.}
\label{layout_2t1r}
\end{figure}

The 2T1R primitive layout is illustrated in fig.\ref{layout_2t1r} with its schematic given in fig.\ref{2T1R} in section 4. As explained in section 4, the 2T1R structure requires the circuit operating up to higher voltages with a pMOS and a nMOS. The layout is highlighted by the net names that correspond to the signals in the schematic. In the layout, the two high voltage transistors are labelled respectively and one memristor placed in between the transistors. The memristor area highlighted by the blue rectangular is magnified. The layout is too large to show all the details, so the selected layers displayed are given in the figure. 

Just like in the 1T1R layout, the transistors are split into fingers with equal length. The high voltage pMOS has six fingers, while the nMOS is split into two. Each two fingers share one drain terminal for pMOS whereas nMOS has a shared source terminal. There are three rows of metal via M1-M2 on the pMOS. The three vias on the middle row are connected together to the memristor and the rest vias are linked through metal 2 layer. In the close look to the memristor connection, the memristor has its top electrode connected to metal M2-M4 vias. These vias are further linked to metal 2 route horizontally extends to both transistors. The route reaches the transistors through M1-M2 via. The bottom layer remains its only connection to metal 4 layer, however, with a wider wire placing among the length of this whole cell. This is because the signal applied to the bottom electrode is a powered signal. 

It can be noticed that all port signals are placed to the edge of the cell with lengths equal to the cell width or length. This is designed on purpose considering to implement the cells in an array, section \ref{Memristor array layout} will demonstrate more in details. The layout still follow the CMOS routing strategies for creating a professional design. For instance, the metal layers are placed in one direction as possible. Odd layer wires are in vertical direction, whereas even layers are routed horizontally. Meanwhile, the ground signal is been placed at the lowest metal layer for minimising noises.
 
More importantly, the layout cells shown above are for illustrative purposes, more compact layout are also possible by shrinking the memristor sizes and keep distances at minimum.
 
\subsection{Memristor array layout} \label{Memristor array layout}
Memristor takes advantages for compact layout cells since it has a simple structure. It has been widely used in array structures among the field such as image processing, neuromorphic systems and other in-memory computing \cite{Li_2017,Yao_2020,Wan_2020}. It is possible to build high density memories by using the memristor array. A crossbar array typically consists of bit-lines and word-lines controlling the unit memory cell . Each of the cross point has a memristive device. This configuration maximises the memory area density. 

When designing an array with memristor, one should take care of sneak current path issue from design point of view. As from layout aspect, main focus has been placed on the line resistance challenge that may occur when a large-scale array is designed. Typically, the wire resistance for a small array can be negligible. However, when it comes to scaling up the array, the wire resistance will result in a considerable voltage drop leading to a signal delivered lower than expected at the memristive devices. High wire resistance blocks the devices from receiving sufficient voltages for operation. It breaks the functionality of the circuit, making the design potentially unusable in the case where not enough voltage can be applied across the memristor device, resulting in no suitable change in its resistance. Line resistance can never be eliminated but it can be kept at a minimum to reduce its negative impact to the circuit. At the same time, we also desire a compact memory cell which allows us to fit more devices in the same area.

Therefore, parasitic analysis is non-avoidable before implementing an array. For a reliable and high density crossbar array, it is suggested to use wider wire width as parasitic resistance usually decreases at a lower rate than capacitance. A trade-off should be considered between the wire width and the size of the layout. The wider the line width, the smaller the wire resistance. At the same time, the area of the layout increases as an increase in line width. For a small memory pitch, the limitation becomes the cell size itself. Moreover, for more complicated design, additional resistance also exist for the array with selectors when the switching components such as transmission gates are included. The introduce of transistors further weaken voltages to the memristive devices. Additionally it is important to look at the digital signals; as the RC constant will affect the minimum pulse width at which the line can be driven.

\begin{figure}[ht]
\centering
\includegraphics [width=15.6cm]{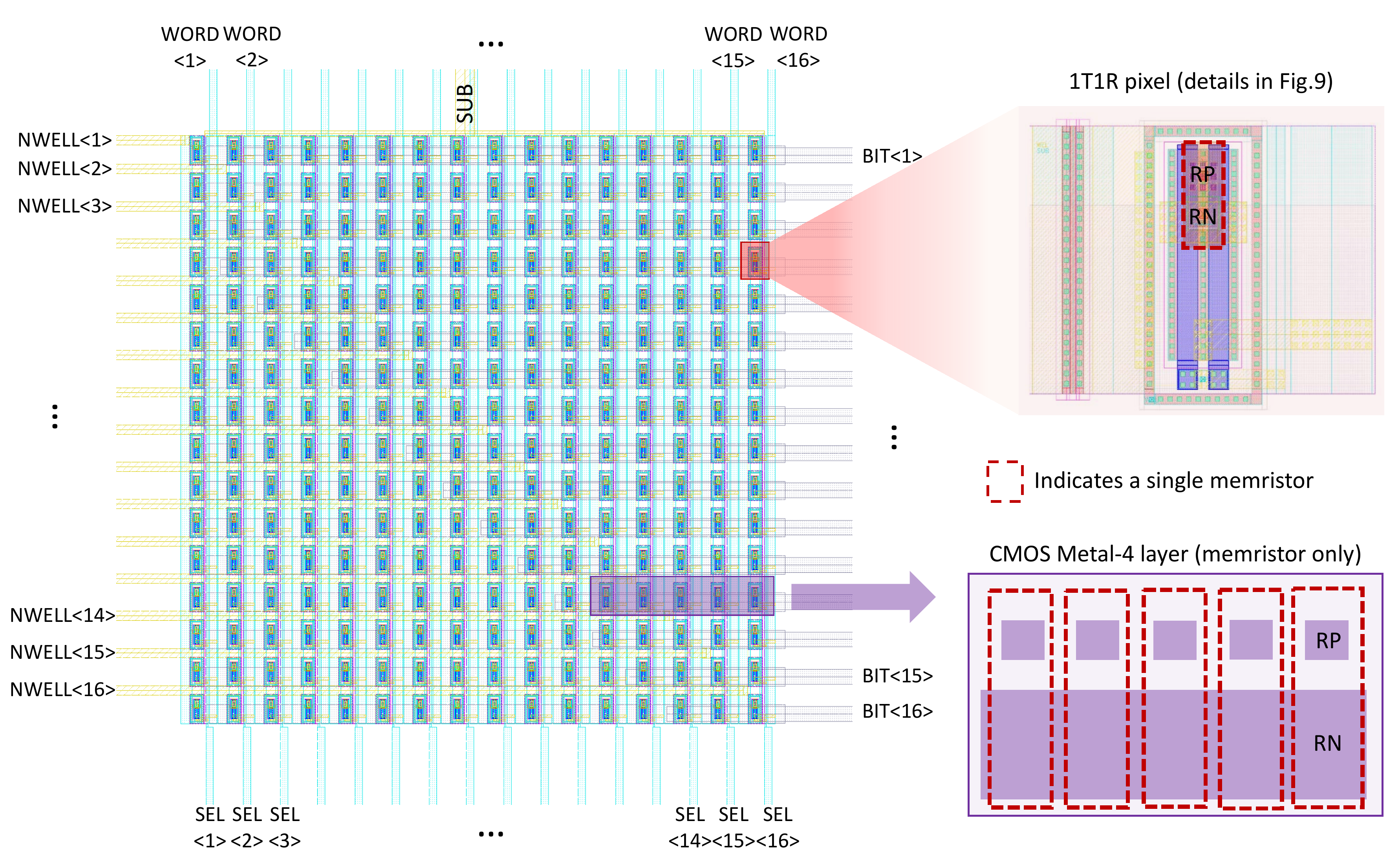}
\caption{Memristor array in a 16 $\times$ 16 1T1R crossbar structure with standard cells shown in Fig.\ref{layout_1t1r}. The  corresponding schematic net names are labelled on the layout. Memristive devices are tied horizontally naming the port BIT$<$1:16$>$ whereas the remaining signals (WORD$<$1:16$>$, SEL$<$1:16$>$, NWELL$<$1:16$>$) are controlled column-wise. The array substrate is connected as a whole by SUB.}
\label{layout_1t1r_array}
\end{figure} 

The 1T1R and 2T1R primitive layout cells demonstrated in Section \ref{1t1r2t1r} can be used into crossbar structures whose maximum array size depends on the line resistance. The overall line resistances can be approximated by measuring the width and length of the corresponding lines. These are then divided to find the number of squares in each line, which is then multiplied by the resistance value in the technology documentation provided by the foundry (usually in Ohms per square).  In our examples, for the 1T1R primitive cell, the analogue signals which are routed with wide widths have a resistance less than 0.25 Ohms. The digital signals have a line resistance slightly larger but tolerable value of about 1 Ohms. Whether these are good enough or not depends on the application for which these memory cells are used.

Fig \ref{layout_1t1r_array} shows a 16 $\times$ 16 array designed using 1T1R standard cell given in fig \ref{layout_1t1r}. The wires at the boundary extend to the signal pads, which are not shown in the figure. The gates (SEL$<$1:16$>$) of the array are connected vertically where each column has one signal to the pads at the bottom.
The n-wells (NWELL$<$1:16$>$) for each cell are also controlled column-wised to the pads at the left side. They are built to control the bulk voltage for every column (discussed in section 4). The top plates of the memristive devices (BIT$<$1:16$>$) in each row are routed together and accessed by the pads located at the right side. The remaining top pads provide the external connections to the pMOS (WORD$<$1:16$>$) for all columns. An additional wire is routed for the array substrates (SUB). There are several ways of improving the standard cell layout for this array, including placing the substrate connection under the n-well. Optimising the layout of the cell is important but not within the scope of this work.

In array layout, the standard cells can be carefully designed with special techniques as used in other arrays found in image sensors and SRAM or DRAM memories. As noticed on the 2T1R layout cell previously in fig.\ref{layout_2t1r}, the port signals are routed till the edge of the cell boundary covering the lengths or the widths of their standard cell. In this way, the implementation of a array is achieved by simply aligning the boundaries of each cell. The wires are auto-connected since they are at the edge if handled by care. This alignment brings convenience to routing an array and keeps the standard cells dense and tight. 

\subsection{Exporting layout to GDSII} \label{Exporting layout to GDSII}

The GDSII stream format is a file format which is the \emph{de facto} industry standard for data exchange of integrated circuit layouts. It is a binary file containing a database of planar objects including geometric shapes, text labels, etc in a hierarchical form. This file alone can be used to transfer designs between different tools or to create masks used in the fabrication process.
 
Designs that are created using the Cadence Design Systems Virtuoso tool suite are typically exported to GDSII for submission to CMOS foundries. 


However, to create a GDSII file containing custom design layers, e.g. for back-end RRAM processing, the designer needs some further insight into the GDSII format itself. The GDSII file format does not use layer names but instead, geometry exists on a numbered layer and datatype. Typically, the layer number and datatype can be in the range 0-255. There is therefore the need to translate the native design layers/purposes to GDSII layer numbers/data types. This is typically achieved using a ``layer mapping table’’ and can be used for both importing and exporting GDS files. For CMOS layers, the default setting specifies this through the technology file that is internally linked to the design kit, using the process listed above. For custom post processing layers, this needs to be specified in a custom layer mapping table. This can be achieved in one of two ways: (1) by extending the technology file using the graphical user interface, or (2) manually through a plain text file. Most CMOS technologies provide both a technology file and layer mapping table, so it is also possible to extend the standard layer mapping table that is provided by the foundry.

However as extending the technology file, means the generated GDSII file will contain both the CMOS design layers and post-processing layers, this may be undesirable if submitting the CMOS design to a foundry. This may be desirable should we wish to contain the entire design (CMOS + RRAM layers) within a single file. The description below therefore focuses on a manual layer mapping description through a plain text file. An example is shown below. Key features are as follows:
\\
\\
\\
\\
\begin{itemize}
\item The Cadence layer purposes are listed in the LSW (layer window). Examples:  dg = drawing, d1=drawing1, pn=pin.
\item Layers that are not listed in the layer mapping table are not exported/imported.
\item Stream layer numbers must be integers between 0 and 255. Numbers 1 through 127 are user defined, and 128 to 255 are reserved for system definition.
\item Generally, only Stream data type 0 should be used for drawing layers.
\end{itemize}
\vspace{1mm}
\begin{Verbatim}[frame=single, label=\large\textrm{{postprocesslayer.map}}, framesep=3mm, fontfamily=tt]
#Cadence layer    Cadence layer    Stream layer    Stream data 
#name             purpose          number          type 
# 
via2rram          drawing          1               0 
electrode1        drawing          2               0 
foxide            drawing          3               0
Electrode2        drawing          4               0
\end{Verbatim}

Finally to export a GDSII stream using the custom layer mapping table, this needs to be specified in the File/Export/Stream form 
i.e. if a custom layer mapping file is specified this is used, otherwise if left blank, the technology file is used.

\section{Discussion and Conclusion}

As we have seen, design with RRAM introduces a unique set of challenges ranging from the definition of the device model, to its incorporation into the CAD toolchain, the actual schematic-level design and all the way to the layout. In this guide we have presented a basic procedure that can take the designer through the entire flow from a CAD tool, to layout design. This is an essential first step before RRAM devices can start entering the mainstream, allowing the community to start embedding RRAM into CMOS and beginning the long effort required for characterising the technology in full and building a solid foundation for eventual incorporation of the technology into the standard toolkit available to the engineer.

Next steps towards the incorporation of RRAM into standard CMOS would include:

\begin{itemize}
    \item Upgrading current RRAM models with e.g. variability data for Monte Carlo analysis and a parasitics model, among others.
    \item Creating a parametrisable cell (p-cell) for schematic design, much akin to the corresponding transistor models.
    \item Building a design rule check (DRC) deck so that layout tools can automatically check the correctness and manufacturability of RRAM device layouts.
    \item Upgrading the LVS deck for automatically recognising RRAM devices.
    \item Providing a small, basic library of fundamental designs (e.g. 1T1R) ready for use by designers, much like logic gates are provided for typical commercial CMOS technologies.
    \item Creating macros for generating e.g. memory blocks using RRAM devices using hardware description languages (HDLs).
\end{itemize} 

These capabilities are expected to slowly mature over time as RRAM technology is tightly integrated with CMOS becoming an increasingly standardised and well-supported part of the CMOS fabric.

\bibliography{Bibliography/bibliography.bib}
\bibliographystyle{unsrtnat}

\newpage
\appendix
\section{Verilog-A Model - }

\subsection{Verilog-A memristor model using exponential fitting}
The Verilog-A memristor model proposed in \cite{Messaris_2018} utilises exponential fitting which  have been shown in Eq.\ref{eq:model_ifunc_app}-\ref{eq:model_analytical_app}. Since the main derivation concept is the same as the one in section 3, we will only go through the some differences in terms of equations, parameter and processing step of this model.\par
The applied DEA set and derived analytically equation shows in the following:
\begin{align}
  \label{eq:model_ifunc_app}
  i(R, v) &= \left\{\begin{array}{ll}
    a_p(1/R)\sinh{(b_p v)}  & v \ge 0 \\
    a_n(1/R)\sinh{(b_n v)}  & v < 0
  \end{array}\right. \\
    \dfrac{dR}{dt} &= g(R, v) = s(v) \cdot f(R, v)
\end{align}
with $s(v)$ being the switching sensitivity function
\begin{equation}
  \label{eq:model_sfunc_app}
  s(v) = \left\{\begin{array}{ll}
      A_p(-1+\exp{(t_p|v|)}) & v > 0 \\
      A_n(-1+\exp{(t_n|v|)}) & v < 0 \\
    0 & \text{otherwise}
  \end{array}\right.
\end{equation}
$f(R,v)$ the window function
\begin{equation}
  \label{eq:model_ffunc_app}
  f(R,v) = \left\{
    \begin{array}{ll}
      -1+\exp{\left[ \eta k_p(r_p(v) - R) \right]} & R < \eta r_p(v) \quad v > 0 \\
      -1+\exp{\left[ \eta k_n(R - r_n(v)) \right]} & R > \eta r_p(v) \quad v < 0 \\
      \phantom{-}0 & \text{otherwise}
    \end{array}
  \right.
\end{equation}
and convert the DAE set to RS time-response equations analytically under constant bias voltage
\begin{equation}
  \label{eq:model_analytical_app}
  R(t)|_{V_b} = \left\{
    \begin{aligned}
        \frac{ln(e^{\eta k_pr_p(V_b)}+e^{-\eta k_ps_p(V_b)t}\times (e^{\eta k_pR_0}-e^{\eta k_pr_p(V_b)}))}{k_p} \quad & for\quad V_b > 0 \& R < \eta r_p(V_b) \\
      \frac{ln(e^{-\eta k_nR_0+\eta k_ns_n(V_b)t})-e^{-\eta k_nr_n(V_b)}\times (-1+e^{\eta k_ns_n(V_b)t})}{k_n} & for\quad V_b < 0 \& R > \eta r_n(V_b) \\
      R_0 \quad \quad \quad \quad \quad \quad \quad \quad \quad \quad \quad \quad \quad & \text{otherwise}
    \end{aligned}
  \right.
\end{equation}


With the aid of equations described in \cite{Messaris_2018}, the Verilog-A memristor model (the in-house fabricated $Pt/TiOx/Pt$ device under $10-17 k\Omega$ $RS$ range) was implemented as presented in Listing 2. Looking into the `window function', it can be found that the positive boundary is fixed to $16.719k\Omega$, which means that the apply voltage above $0.5V$ can reach this $RS$.
\begin{figure}[h!]
\centering
\includegraphics [width=15cm]{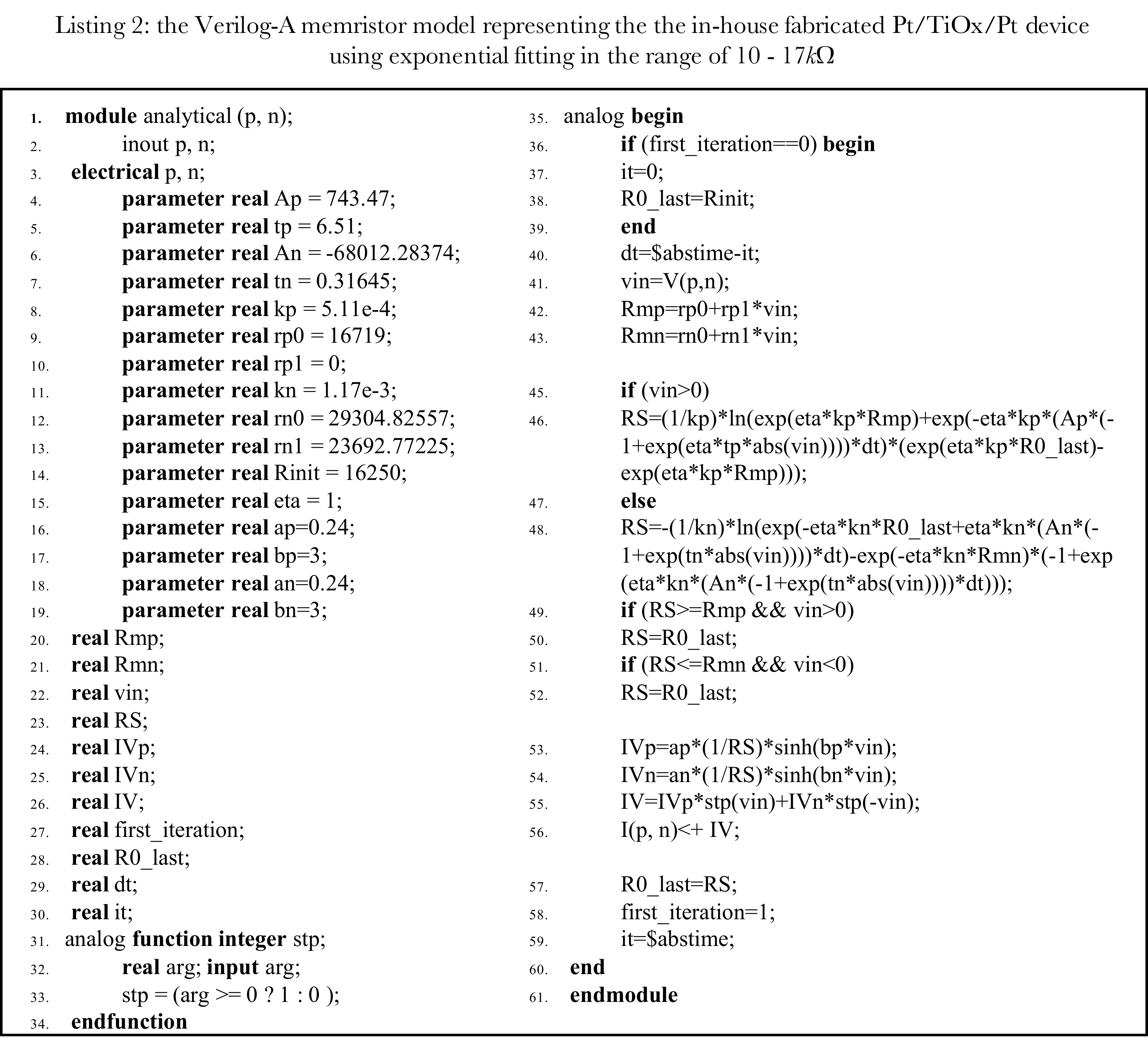}
\label{Verilog-ACode}
\end{figure}

This is the data-driven model that we obtained by applying multiple voltages on the device based on the parameter extraction algorithm, and details can be found in \cite{Messaris_2018}. In this stage, we present two $RS$ ranges that has been proved to fit our physical model in low RMS errors. Considering that the $RS$ is mainly depended on the applied voltage under the same initial resistive state, both suggested bias voltages and $RS$ ranges will be given. The suggested applied voltage is around between $|1.2V|$ and $|1.5V|$ in the $RS$ range of $[4.5k\Omega , 6.0k\Omega ]$. While for the range of $[10k\Omega , 17k\Omega ]$, users are suggested to apply voltage between $|0.6V|$ and $|0.8V|$. But in this case, we use the model in Listing 2 for demonstration to exploit impacts from different types of stimulus, including pulse number, pulse width and amplitude.
Fitting parameters for two operational $RS$ range using exponential show in Table \ref{ParameterTable_app}.

\begin{table}[!h]
    \centering
    \caption{Parameter values that fit the $Pt/TiOx/Pt$ memristor in two $RS$ ranges.}
    \begin{tabular}{ c c c }
 \hline 
 Parameters & $Pt/TiOx/Pt$ ($4.5-6.0k\Omega$) & $Pt/TiOx/Pt$ ($10-17k\Omega$)\\
    \hline
    $A_p$ & 0.12 & 743.47 \\ 
    $A_n$ & -79.03 & $-6.8\times10^4$ \\ 
    $t_p$ & 0.59 & 6.51 \\
    $t_n$ & 1.12 & 0.31 \\
    $k_p$ & $8.10\times 10^{-3}$ & $5.11\times 10^{-4}$\\
    $k_n$ & $9.43\times 10^{-3}$ & $1.17\times 10^{-3}$\\ 
    \hline
    $r_{p0}$ & 3085 & $16.71\times 10^3$\\
    $r_{p1}$ & 1862 & 0\\
    $r_{p2}$ & 0 & 0\\ 
    \hline
    $r_{n0}$ & 5193 & $29.30\times 10^3$\\
    $r_{n1}$ & 378 & $23.69\times 10^3$\\
    $r_{n2}$ & 0 & 0\\ 
    \hline
    $a_{p,n}$ & 0.24 & 0.24\\
    $b_{p,n}$ & 2.81 & 2.81\\
 \hline
\end{tabular}

    \label{ParameterTable_app}
\end{table}


\subsection{Calibration of exponential verilog-A model}
This section aims at providing users with the operational range and performance of the proposed model, as well as how device parameters affect the above performance.
We program the proposed model by modulating number of pulse (in A.2.1), pulse width (in A.2.2) and amplitude (in A.2.3) in order to explore the both qualitative and quantitative impacts on model. To keep the consistency, all simulations will start from the same initial $RS$ as baseline. Recommendation will be given after evaluating above modulations. 

\subsubsection{Number of pulse modulation}
In this part, positive bias voltages from $0.6V$ to $0.8V$ are employed on the Verilog-A memristor model. The reason we choose the positive voltage is that in our model the boundary of upper limit is fixed and will not be affected by the bias voltage (refer to Listing 1, lines 9, 10 and 42). Thus, we can explore the phenomenon when we applied unlimited number of pulse on the model. The simulation will be setup as applying three level of amplitude voltages on the device and run the simulation till the $RS$ become saturated.\par

\begin{figure}[h!]
\centering
\includegraphics [width=10cm]{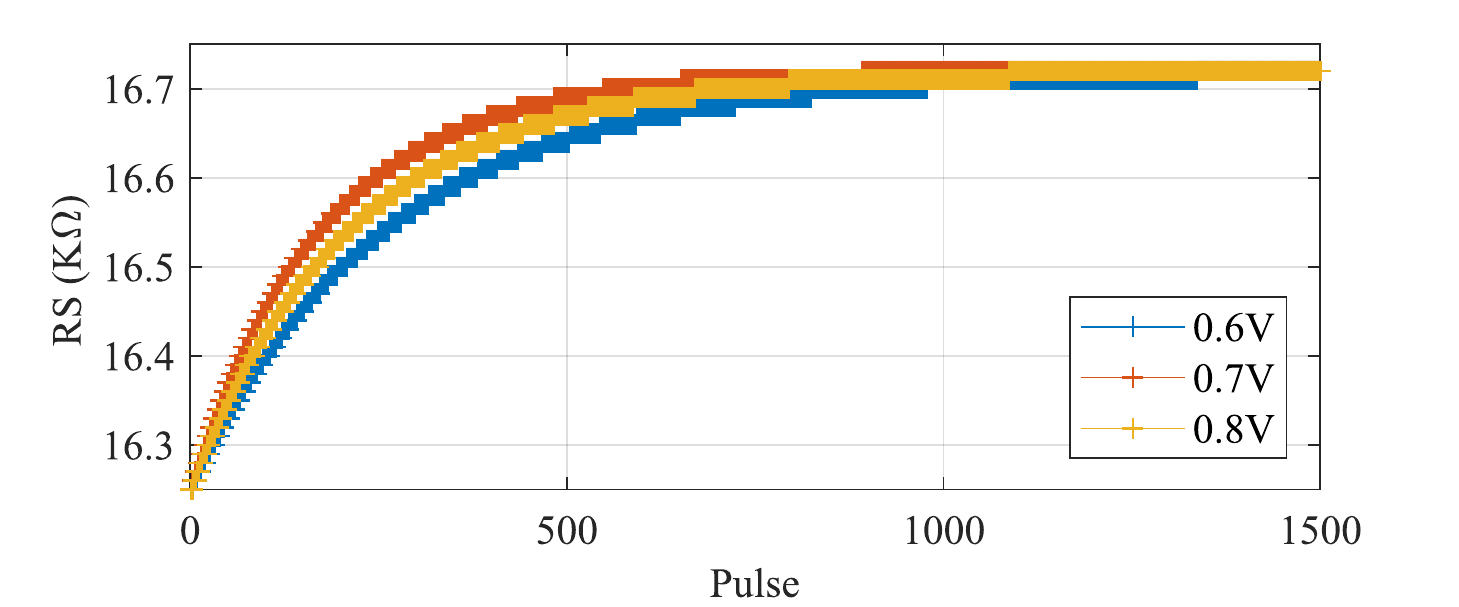}
\caption{Verilog-A memristor model response based on the number of applied pulses. The device is provoked by 1500 pulses with different voltage starting from initial $RS=16.25k\Omega$, and it eventually saturated at $RS=16.71k\Omega$. As the pulses keep provoking, changing rate of $RS$ slows down and it gradually saturated. Characterisation routine parameters based on the stimulus in Figure \ref{InputSignals}: $t_{w,\Delta R}=100\mu s$, $t_{w,iv}=1.1ms$, $V_b=0.6/0.7/0.8V$, and $V_{read}=0.5V$}
\label{DiffPulseApp}
\end{figure}

\subsubsection{Pulse width modulation}
Exploration of pulse width effects on programming the memristive devices will be given in this subsection, where we generate pulses at the same bias voltage, $|0.8V|$, in three types of width: $1\mu s$, $10\mu s$ and $100\mu s$. In Figure \ref{DiffWidthTrans}, three trains with a number of 500 pulses are generated to discover the changing process and the eventual result of $RS$, which presents in Figure \ref{DiffWidth}. Besides, negative voltage are employed on the device to flush it back to initial $RS$, where helps us prove that the modulation in positive voltage can be applied on negative one. \par
Figure \ref{DiffWidth} presents pulse width modulation of positive voltage, the stable $RS$ programmed by $100\mu s$ pulse cannot be pinpointed especially in the first 200 pulses, compared with two other pulse width stimulus. It indicates that pulse with smaller width can be applied to slow down the changing rate, which contributes to programming the device to a specific resistive state in application.

\begin{figure}[h!]
\centering
\includegraphics [width=9.5cm]{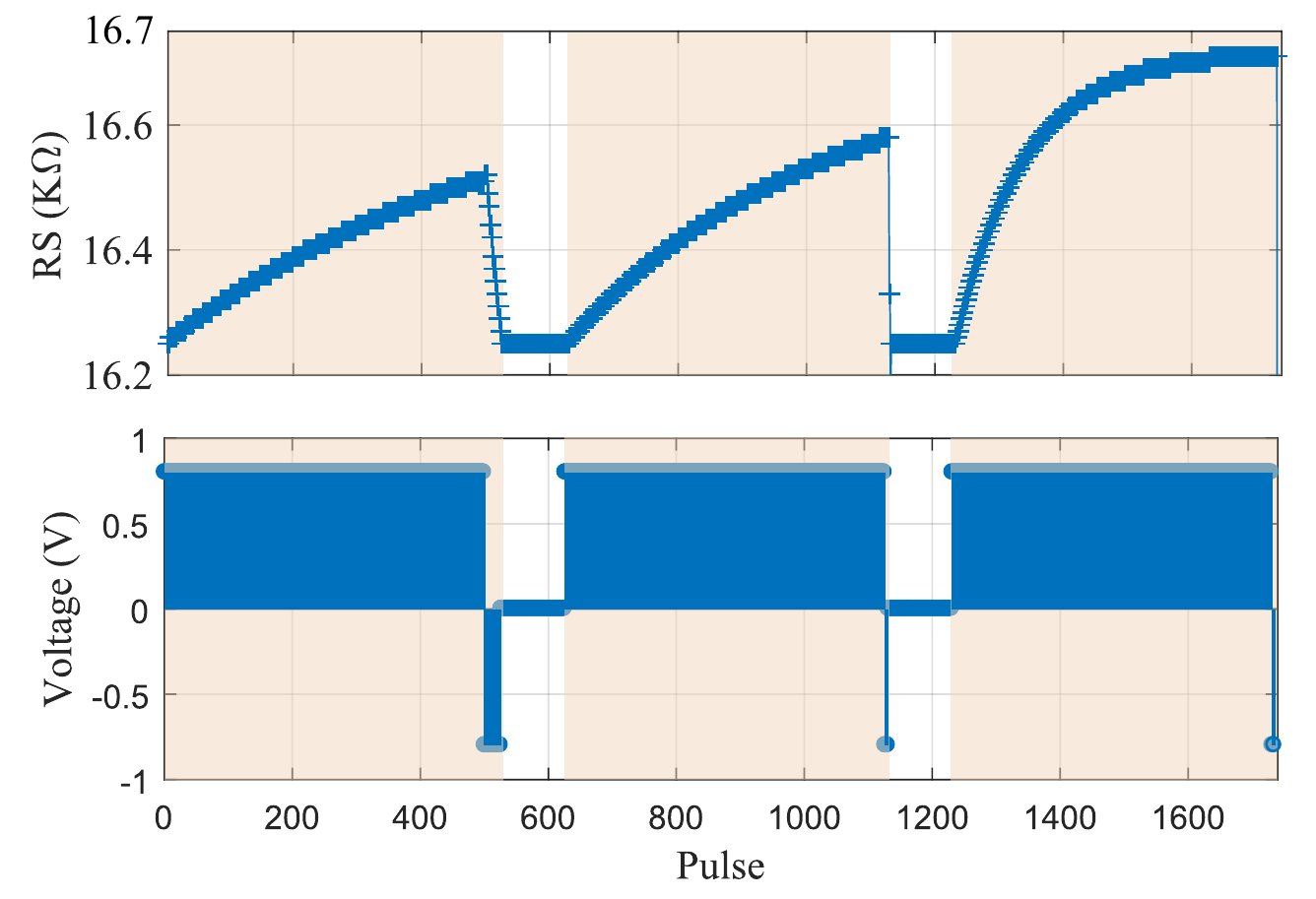}
\caption{Programming the device with three types of duration pulses. Bottom trace: Each number of 500 pulses at $0.8V$ with three different duration are employed to modulate device $RS$. In between the measurement, the inverse voltages are applied to flush the device to initial $RS$. Top trace shows the modulation results corresponding to the bottom stimulus with both positive and negative bias voltage. Results with highlight will be shown in Figure \ref{DiffWidthApp} with a clear view of programming memristor to a specific $RS$. Characterisation routine parameters: $t_{w,\Delta R}=1/10/100\mu s$, $t_{w,iv}=1.001/1.01/1.1ms$, $V_b=|0.8V|$, and $V_{read}=0.5V$.}
\label{DiffWidthTransApp}
\end{figure}

\begin{figure}[h!]
\centering
\includegraphics [width=10.5cm]{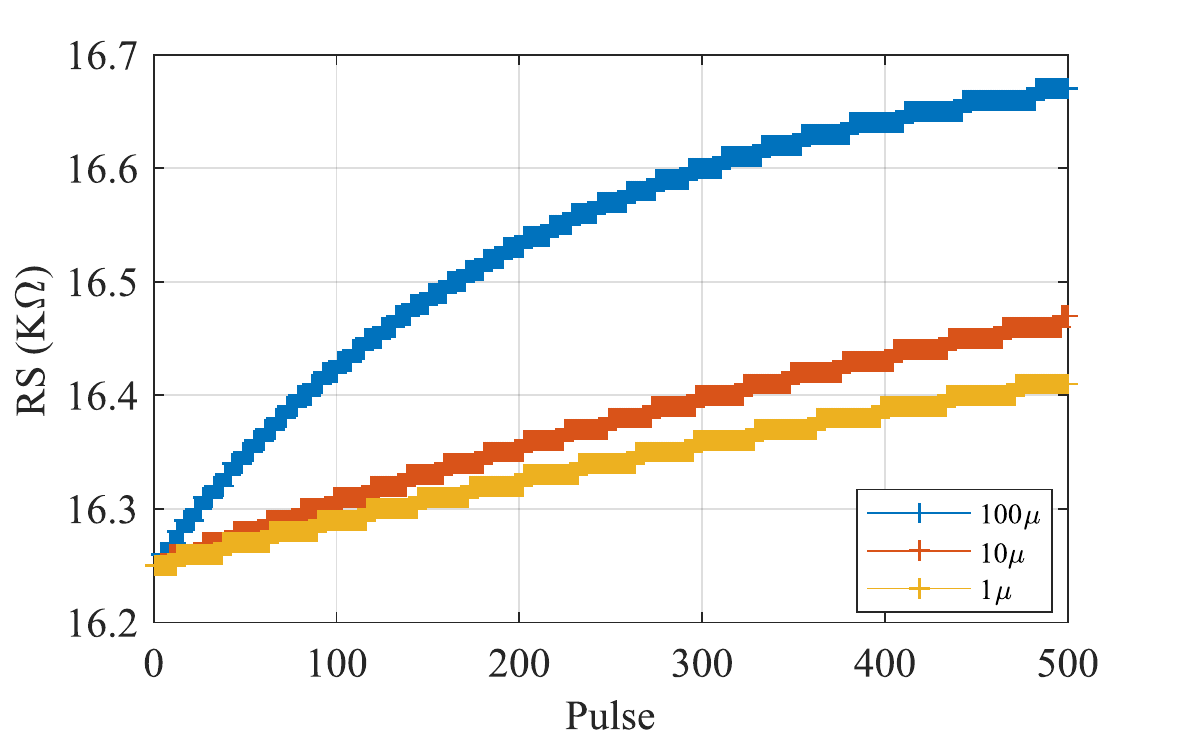}
\caption{Verilog-A memristor model response based on the width of applied pulses. Starting from initial $RS=16.25k\Omega$, resistive states climb in different rates under pulses with different duration. The $100\mu s$ pulses induce that $RS$ increases in a faster speed and generates lots of states within 200 pulses which is less recognisable compared with $1\mu s$ pulses. However, the more stable state (with longer duration) appears as the number of pulse increases among three situations. It illustrates that the shorter duration pulse helps generate specific $RS$ with higher resolution.}
\label{DiffWidthApp}
\end{figure}

\subsubsection{Amplitude and polarity modulation}
\begin{figure}[h!]
\centering
\includegraphics [width=9.5cm]{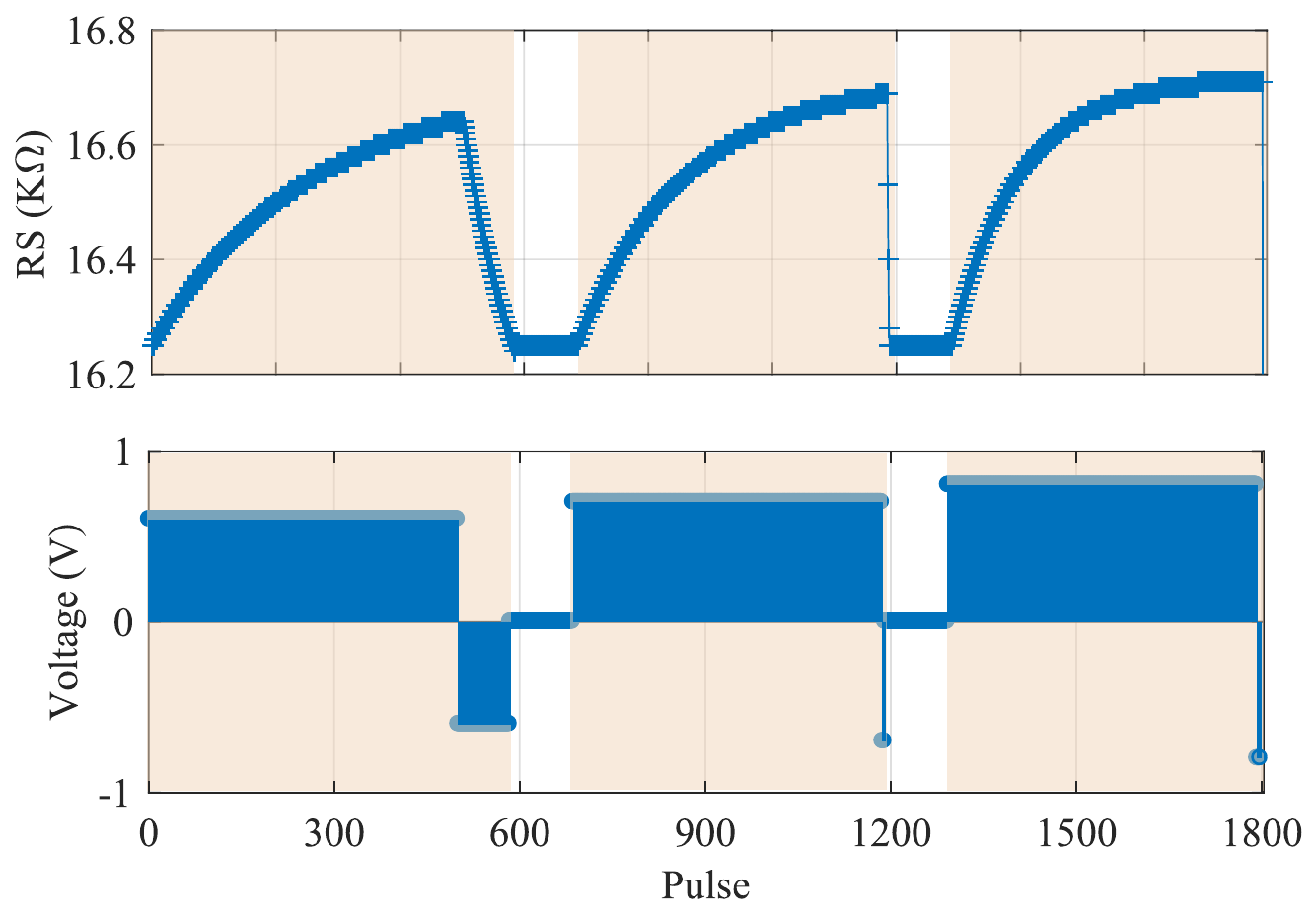}
\caption{Programming the device with three amplitude pulses. Bottom trace: Each number of 500 pulse (duration=$100\mu s$) under incremental bias voltage are employed to modulate device $RS$. In between the measurement, inverse voltages help flush the device back to initial state. The $RS$ modulated by positive bias voltage has been highlighted and will be shown in Figure \ref{DiffAmpApp} for comparison. Characterisation routine parameters: $t_{w,\Delta R}=100\mu s$, $t_{w,iv}=1.1ms$, $V_b=|0.6/0.7/0.8V|$, and $V_{read}=0.5V$.}
\label{DiffAmpTransApp}
\end{figure}

To evaluate the amplitude and polarity modulation, stimulus contains bias voltage from $|0.6V|$ to $|0.8V|$ in two polarities are generated.  A number of 500 pulse trains were applied on the device in order to make comparison in Figure \ref{DiffWidthApp} with fixed pulse number. Besides, negative pulses are also applied on the device to push the device back to the same initial state (in Figure \ref{DiffAmpTransApp}), where builds a comparison of filling the same $RS$ gap by positive and negative bias voltage.\par

\begin{figure}[h!]
\centering
\includegraphics [width=10.5cm]{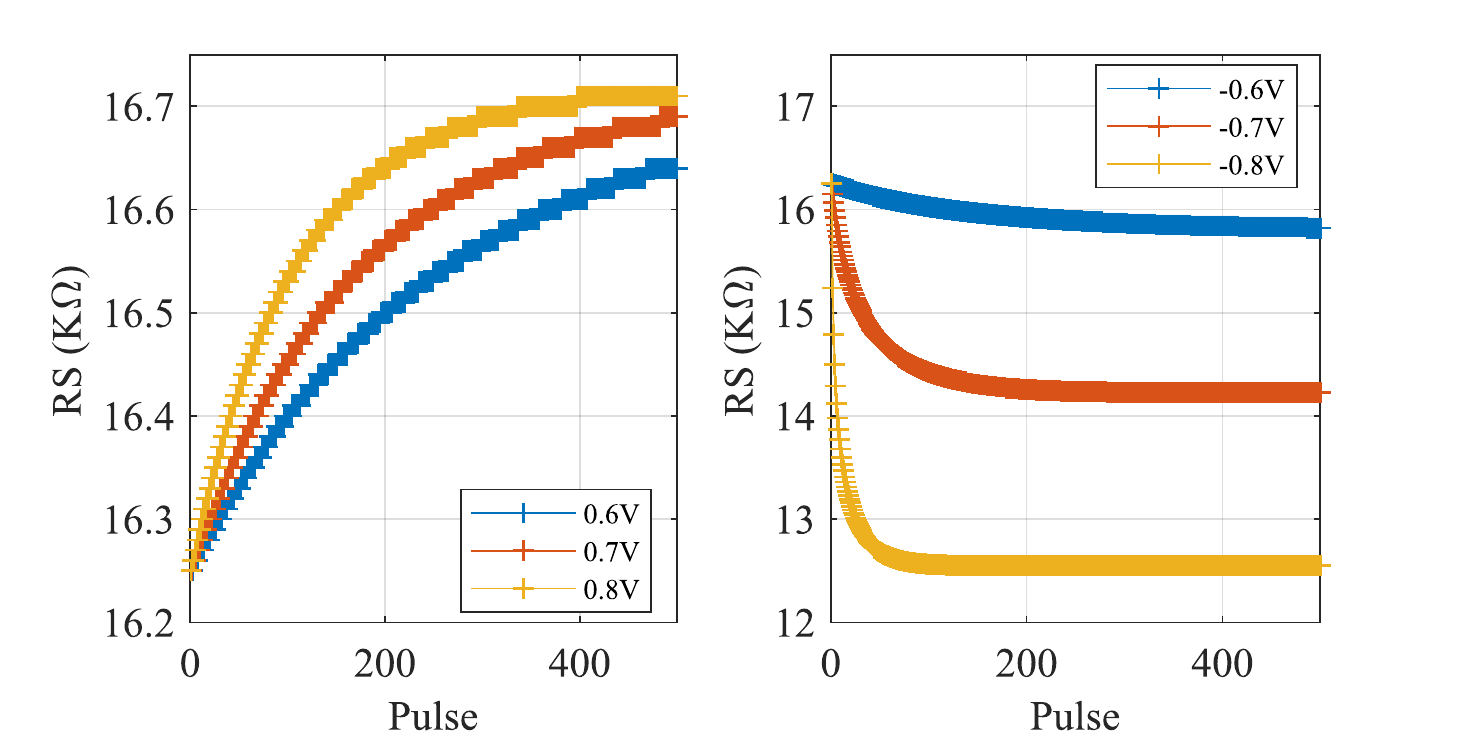}
\caption{Verilog-A memristor model response based on the amplitude of applied pulses. Being stimulated by different amplitude voltage, simulation results from both positive and negative bias voltages present in left and right figure respectively. It can be seen that the higher absolute voltage leads to faster changing rate of $RS$. Combining two sub-figures, positive voltage induces lower changing rate of $RS$, compared with negative voltage. The right figure also indicates that the saturated $RS$ (boundary of lower limit) is dependent on the bias voltage, where a more negative voltage can push $RS$ to a lower state.}
\label{DiffAmpApp}
\end{figure}

Combining above simulation results, some recommendations for application haven been concluded:
\begin{itemize}
    \item These two Verilog-A model are presented in a lower $RS$ comparing with the one in section 3. It can be observed that it needs lower bias voltage to realise switching under the condition of lower initial $RS$. Therefore, it is suggested for user to estimate the suitable bias voltage by applying different amplitudes to a specific $RS$.
    \item Even though these two models have their own $RS$ operation range, the simulation can still completed but $RS$ will not be changed when setting simulation is beyond the `window function'. Since the $RS$ range is limited, users are supposed to monitor the change of $RS$ to prevent $RS$ excess of boundary.
\end{itemize}

\end{document}